\documentclass[prl,reprint,twocolumn,showpacs,superscriptaddress]{revtex4-1}

\usepackage{bm,mathrsfs}
\usepackage{graphicx}
\usepackage{epsfig}
\usepackage{amsmath,bbm}
\usepackage{amsfonts,amssymb}
\usepackage{times}
\usepackage{dsfont}
\usepackage{enumitem}  
\usepackage{comment}
\usepackage[colorlinks,linkcolor=blue,citecolor=blue]{hyperref}

\newcommand{\eqn}[1]{
\begin{eqnarray}
	#1
\end{eqnarray}
}

\begin{document}
\title{Absence versus Presence of Dissipative Quantum Phase Transition in Josephson Junctions} 
\author{Kanta Masuki}
\email{masuki@g.ecc.u-tokyo.ac.jp}
\affiliation{Department of Physics, University of Tokyo, 7-3-1 Hongo, Bunkyo-ku, Tokyo 113-0033, Japan}

\author{Hiroyuki Sudo}
\affiliation{Department of Physics, University of Tokyo, 7-3-1 Hongo, Bunkyo-ku, Tokyo 113-0033, Japan}

\author{Masaki Oshikawa}
\affiliation{Institute for Solid State Physics, University of Tokyo, Kashiwa, Chiba 277-8581, Japan}
\affiliation{Kavli Institute for the Physics and Mathematics of the Universe (WPI),
University of Tokyo, Kashiwa, Chiba 277-8583, Japan}

\author{Yuto Ashida}
\email{ashida@phys.s.u-tokyo.ac.jp}
\affiliation{Department of Physics, University of Tokyo, 7-3-1 Hongo, Bunkyo-ku, Tokyo 113-0033, Japan}
\affiliation{Institute for Physics of Intelligence, University of Tokyo, 7-3-1 Hongo, Tokyo 113-0033, Japan}

\begin{abstract}
Dissipative quantum phase transition has been widely believed to occur in a Josephson junction coupled to a resistor despite a lack of concrete experimental evidence. Here, on the basis of both numerical and analytical nonperturbative renormalization group (RG) analyses, we reveal breakdown of previous perturbative arguments and defy the common wisdom that the transition always occurs at the quantum resistance \(R_{Q} \!=\! h/(4e^2)\). We find that RG flows in nonperturbative regimes induce  nonmonotonic renormalization of the charging energy and lead to a qualitatively different phase diagram, where the insulator phase is strongly suppressed to the deep charge regime (Cooper pair box), while the system is always superconducting in the transmon regime. We identify a previously overlooked dangerously irrelevant term as an origin of the failure of conventional understandings. Our predictions can be tested in recent experiments realizing high-impedance long superconducting waveguides and would provide a solution to the long-standing controversy about the fate of dissipative quantum phase transition in the resistively shunted Josephson junction.
\end{abstract}

\maketitle

Understanding physical properties of quantum systems interacting with environmental degrees of freedom is one of the central problems in quantum many-body physics. A wide variety of intriguing quantum phenomena have been revealed in the last half century; key examples include the Kondo problem in heavy fermion materials or mesoscopic structures \cite{Kondo64,Anderson61,Affleck92,Pustilnik04,LH07}, transport through quantum nanowire systems \cite{Kane92,Furusaki93,Oshikawa06,Bockrath99,Yao99}, and quantum dissipative systems \cite{Caldeira81,*Caldeira83a,*Caldeira83b,Leggett87,Anders07,HUR20082208}. 
One of the most notable predictions among such fundamental problems is the dissipative quantum phase transition (DQPT) occurring in the resistively shunted Josephson junction (RSJ) \cite{Schmidt83,Bulgadaev84,Schon90,NN99,Halperin11,Weiss12,KA19}. Previous studies \cite{Guinea85,Aslangul85,Fisher85,Zaikin87} predicted that the Josephson junction (JJ) at zero temperature remains superconducting below the quantum resistance \(R\!<\!R_Q\!=\!h/(4e^2)\) while it becomes insulator (or precisely normal metal) in \(R\!>\!R_Q\). This result has been obtained by such theoretical methods as   perturbative renormalization group (RG) analysis \cite{Guinea85,Aslangul85,Weiss85,Fisher85,Zaikin87,Averin90,Callan90,RG032,*RG07} and path-integral Monte-Carlo method  \cite{Herrero02,Kimura04,Werner05a,*Werner05b,*Lukyanov07}. While experimental attempts to observe DQPT have been made \cite{Kuzmin91,Yagi97,Penttila99,Penttila01,Liu09}, interpretation of these results has remained a matter of debate  \cite{Penttila01,Herrero02,Murami20,Hakonen21,Murani21}. In particular, a possible absence of  DQPT in the predicted parameter regime has been recently reported \cite{Murami20}. All in all, despite many years of research, a comprehensive understanding of DQPT has yet to be achieved.

\begin{figure}[b]
    \includegraphics[width = 8.6cm]{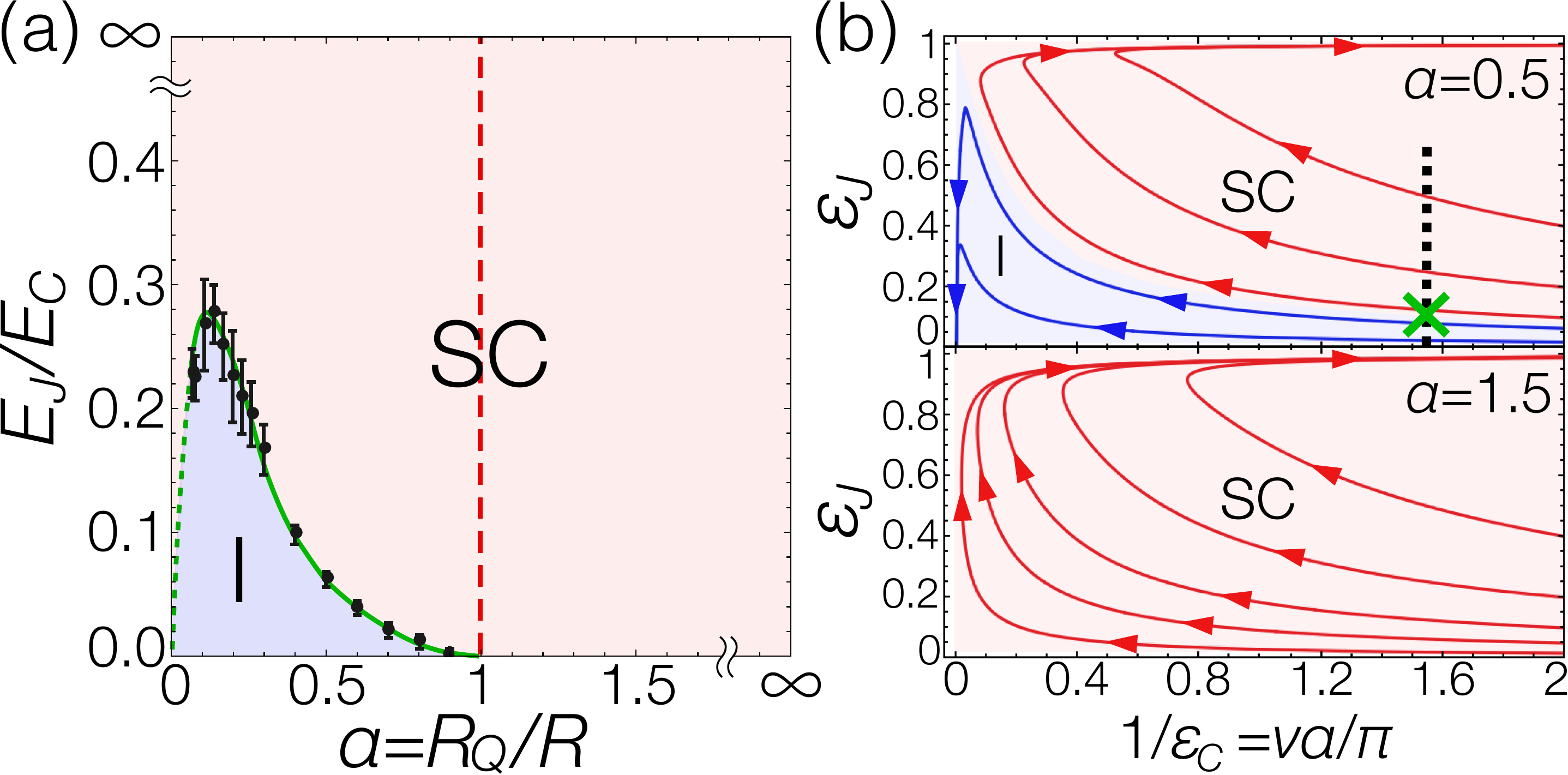}
    \caption{(a) Ground-state phase diagram of RSJ. Green curve indicates the phase boundary determined from NRG, separating the superconducting (SC)  and insulator (I) phases. Red vertical dashed line is the commonly believed boundary. (b) FRG flow diagrams of dimensionless Josephson (charging) energy $\epsilon_{J(C)}$ at different dissipation strengths $\alpha$. At UV scale $\epsilon_{J,C}\!\ll\!1$, Josephson coupling $\epsilon_J$ is always relevant and triggers nonmonotonic renormalization of dangerously irrelevant term $\nu\!\propto\!1/\epsilon_C$. Transition occurs at finite $E_J/E_C$ in $\alpha\!<\!1$ (green cross in top panel), while the system always flows to the SC fixed point in $\alpha\!>\!1$ (bottom panel). Previous perturbative results are reproduced in the limit \(\nu\!\propto\!1/\epsilon_C \!\rightarrow\! 0\).\label{fig1}}
\end{figure}

The aim of this Letter is to fill this gap and provide a solution to the long-standing controversy regarding DQPT. To this end, we  systematically analyze RSJ on the basis of numerical and analytical nonperturbative approaches, namely, numerical renormalization group (NRG) and functional renormalization group (FRG). Surprisingly, both analyses lead to the ground-state phase diagram (Fig.~\ref{fig1}) that is dramatically different from the one expected from the previous arguments. Specifically, the insulator phase is strongly suppressed to the deep charge regime $E_J/E_C\!\ll\!1$ (Cooper pair box) while the system is always superconducting in the transmon regime $E_J/E_C\!\gg\!1$, where $E_J$ is the Josephson coupling and $E_C\!=\!(2e)^2/2C_J$ is the charging energy with the capacitance $C_J$. In particular, as $\alpha\!=\!R_Q/R$ is decreased, our results indicate the reentrant transition from insulating to superconducting phase in $\alpha\!\ll\! 1$ (see also Fig.~\ref{fig4} below). These findings sharply contrast with the common wisdom that the transition should occur at \(R \!=\! R_Q\) for any $E_J/E_C$ (red dashed line in Fig.~\ref{fig1}(a)). 
 
While the conventional understanding at an early stage was made by  perturbative analyses and duality argument, we point out that these previous considerations implicitly discarded a term (which we call the capacitance term $\nu\!\propto\!1/E_C$) that was expected to be irrelevant from dimensional counting~\cite{Fisher85, Affleck01}. We show that this previously overlooked term is actually {\it dangerously irrelevant}, i.e., it can turn into relevant at low-energy scales due to nonperturbative renormalization (Fig.~\ref{fig1}(b)). It is this subtle, yet crucial missing piece that completes our understanding of DQPT and explains the failure of the previous arguments. 

From a broader perspective, small quantum systems interacting with a bosonic bath as studied here are fairly ubiquitous in e.g., electron-phonon  systems and quantum light-matter systems. Our analyses should have a broad range of applications to those systems which are currently the subject of intense research in different fields. Moreover, in view of the fundamental role of JJ in  quantum circuits  \cite{YN99,Blais21,Goldstein13,Leppakangas18,Kaur20,PIV20,Houzet20,BA21}, the present study will also advance our understanding of the interaction between quantum information processors and electromagnetic environments in general.

{\it Model.---} We consider the following RSJ Hamiltonian, in which JJ couples to the environmental degrees of freedom represented as a collection of harmonic oscillators \cite{Caldeira83a}:
\eqn{
  \hat H &=& E_C\left(\hat N-\hat n_r\right)^2\!\!-E_J\cos(\varphi) + \!\!\!\sum_{0<k\leq K}\! \hbar\omega_k\hat a_k^\dagger\hat a_k,\label{ham1}\\
  \hat n_r &=& \frac{\sqrt{\alpha}}{2\pi}\sum_{0<k\leq K} \sqrt{\frac{2\pi}{kL}}(\hat a_k^\dagger+\hat a_k),
}
where $\varphi$ is the JJ phase,  \(\hat N \!=\! -i\partial/\partial\varphi\) is the charge operator, bath frequencies are \(\omega_k \!=\! vk \!=\! vm\pi/L\) with \(m\!=\!1, 2, \cdots M\),  \(K \!=\! M\pi/L\) is the wavenumber cutoff, and \(\hat a_k\) (\(\hat a_k^\dagger\)) is the bosonic annihilation (creation) operator of mode $k$. The constants \(v\) and \(L\) have the dimensions of velocity and length, and \(\alpha = R_Q/R\) is the dimensionless frictional coefficient. We remark that Eq.~\eqref{ham1} takes the same form as in quantum light-matter Hamiltonian under the long-wavelength approximation \cite{CCT89}.  Below we aim to extract its physical properties in the wideband condition $E_{J,C}\!\ll\!\hbar W$ and thermodynamic limit \(L\!\rightarrow\! \infty\), where we denote the frequency cutoff as \(W \!=\! vK\).

We first diagonalize the quadratic part, \(E_C\hat n_r^2 \!+\! \sum_k\hbar\omega_k\hat a_k^\dagger\hat a_k\), via the Bogoliubov transformation and rewrite the Hamiltonian~\eqref{ham1} as (see e.g., Ref.~\cite{Affleck01})
\begin{align}
  \hat H =& E_C\hat N^2 - E_J\cos(\varphi) \nonumber\\
  &-\hat N \sum_{0<k\leq K}\hbar g_k(\hat b_k+\hat b_k^\dagger) + \sum_{0<k\leq K}\hbar \omega_k\hat b_k^\dagger\hat b_k,\label{ham2}\\
 g_k =& \sqrt{\frac{2\pi v}{\alpha L}\frac{\omega_k}{1+\left(\frac{\nu\omega_k}{W}\right)^2}},\;\; \nu \equiv \frac{\pi}{\alpha\epsilon_C}, \;\;\epsilon_C = \frac{E_C}{\hbar W},\label{capacitive coupling}
\end{align}
where we introduce the squeezed annihilation (creation) operators \(\hat b_k\) ($\hat b_k^\dagger$). The Hamiltonian~\eqref{ham2} can also be derived from a microscopic model of JJ shunted by a transmission line with impedance \(R\), length \(L\), and propagation speed \(v\) \cite{SM1}. A salient feature is that the capacitive coupling \(g_k\) acquires suppression at frequencies higher than  \(W/\nu \!= \!\alpha E_C/(\pi\hbar)\) \cite{GMF17,MM17,Parra-Rodriguez18}. This natural cutoff frequency, $\alpha E_C/(\pi\hbar)$, depends only on the model parameters and our results are independent of a choice of $W$ as long as the wideband condition, \(W\!\gg\!\alpha E_C/(\pi\hbar)\), is satisfied. 

To perform the NRG analysis  \cite{Wilson75}, we next use a unitary transformation \(\hat U \!=\! \exp(-i\hat N\hat \Xi)\) with \(\hat \Xi\!=\!i\sum_k \frac{g_k}{\omega_k}(\hat b_k^\dagger-\hat b_k)\) \cite{YA21,*ashida21}. Introducing the field operators \(\hat \phi(x)\) and \( \hat \pi(x)\), 
\eqn{
  \hat \phi(x) &=& \sqrt{\alpha}\varphi + \sum_{0<k\leq K}\sqrt{\frac{2\pi}{kL}}i(\hat b_k-\hat b_k^\dagger)\cos(kx),\\
  \hat \pi(x) &=& \sum_{0<k\leq K}\sqrt{\frac{2\pi k}{L}}(\hat b_k+\hat b_k^\dagger)\sin(kx),
}
we obtain the transformed Hamiltonian \(\hat H_U \equiv \hat U^\dagger \hat H\hat U\),
\begin{align}
  \hat H_U =& -E_J\cos\left(\frac{1}{\sqrt{\alpha}}\int_0^L dx\hat \phi(x)f_\nu(x)\right)+\hat{H}_{\rm TLL},\label{ham3}
\end{align}
where $\hat{H}_{\rm TLL}$ is the Tomonaga-Luttinger liquid Hamiltonian and $f_\nu(x)$ is the function that exponentially vanishes on the length scale $\nu/K\!=\!\pi\hbar v/(\alpha E_C)$ as follows:
\begin{align}
 \hat{H}_{\rm TLL}=&\frac{\hbar v}{4\pi}\int_0^L dx\left[\left(\partial_x\hat\phi(x)\right)^2+\hat \pi(x)^2\right],\\
  f_\nu(x) =& \frac{2}{\pi}\int_0^{K} dk\frac{\cos(kx)}{\sqrt{1+(\nu k/K)^2}}.\label{f_eta}
\end{align}
To derive Eq.~\eqref{ham3}, we use the sum rule, \(\sum_k \hbar g_k^2/\omega_k \!=\! E_C\), which can be shown for a general light-matter-type Hamiltonian \cite{YA21,*ashida21}.
 The new frame~\eqref{ham3} gives a proper basis to extend Wilson's NRG approach to RSJ \cite{footnote_kinetic, SM1}.

{\it Benchmark results: the boundary sine-Gordon model.---} 
Before analyzing the exact RSJ Hamiltonian~\eqref{ham3}, we start from benchmarking our NRG analysis for the boundary sine-Gordon (bsG) model \cite{Guinea85,Fisher85,Kane92,Furusaki93,FP95,Affleck01}:
\begin{align}
  \hat H_{\rm bsG} = &-E_J\cos\left(\frac{\hat \phi(0)}{\sqrt{\alpha}}\right)+\hat{H}_{\rm TLL},\label{ham4}  
\end{align}
which can be obtained by taking the limit $\nu\!\to\!0$ in Eq.~\eqref{ham3}.
Its ground-state properties are well understood from the perturbative analysis, which predicts the transition at $\alpha\!=\!1$. When \(\alpha \!>\!1\), the Josephson coupling \(E_J\) is relevant and, in the original frame~\eqref{ham1}, leads to the phase localization around  \(\varphi\!\sim\! 2\pi\mathbb{Z}\). In other words, the ground state is phase-coherent and superconducting. Conversely, when \(\alpha\!<\!1\), the Josephson energy $E_J$ renormalizes to zero and the charge becomes localized, i.e., the system is insulating.

\begin{figure}[tbh]
    \includegraphics[width=8.6cm]{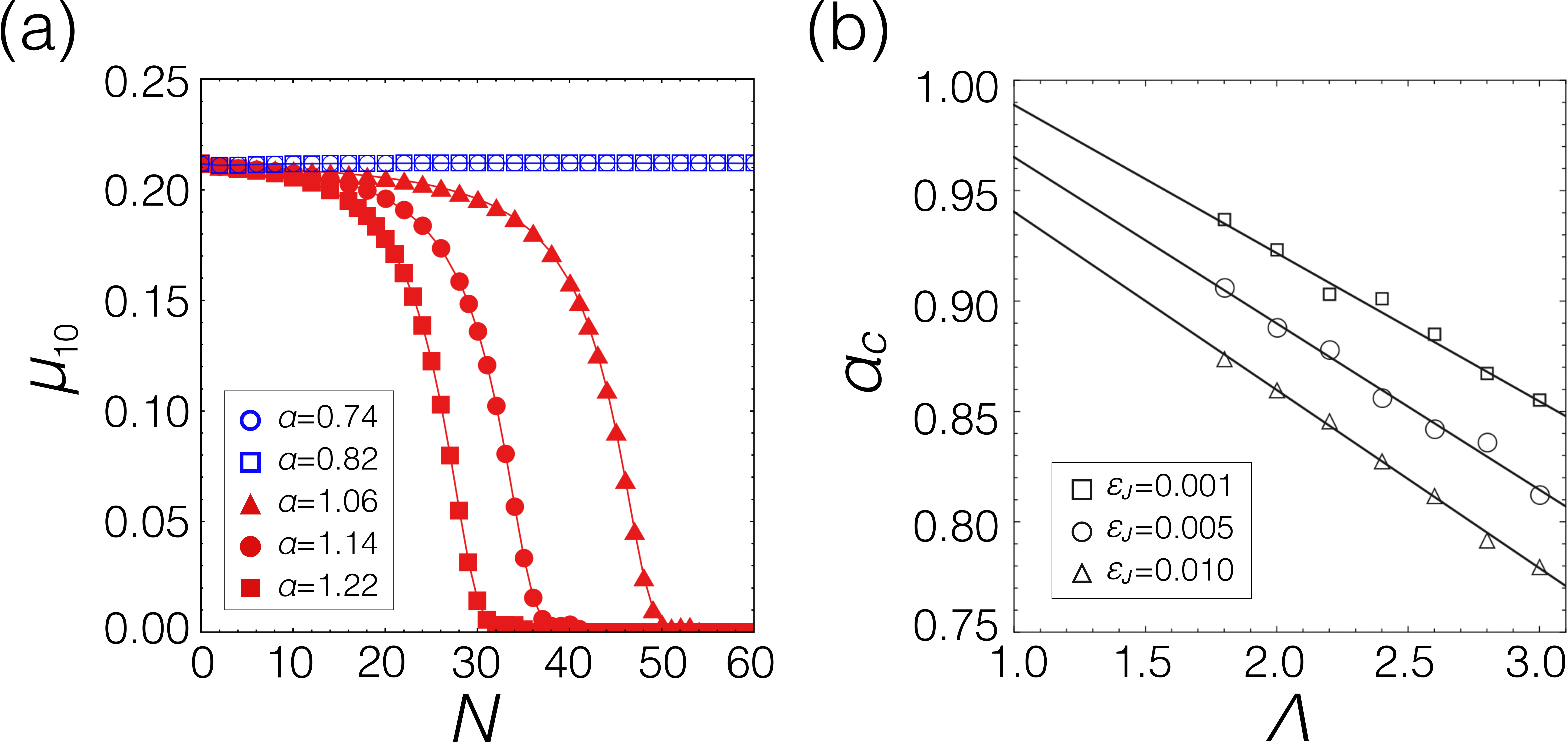}
    \caption{NRG benchmark results in the bsG model~\eqref{ham4}. (a) Flows of the mobility \(\mu_{10}\) plotted against the number of RG steps $N$. In the SC phase $\alpha\!>\!\alpha_c$, the mobility flows to zero (red curves), while it remains nonzero in the insulator phase $\alpha\!<\!\alpha_c$ (blue curves). Parameters are \(\epsilon_J \!=\! 0.001\) and \(\Lambda \!=\! 2.0\). (b) Extrapolations of the critical value \(\alpha_c\) to the Wilson parameter \(\Lambda\!\to\!1\). The scaling limit \(\epsilon_J\!=\!E_J/\hbar W \!\rightarrow \!0\) leads to the transition point \(\alpha_c = 0.99(2)\) which agrees with the analytical value $\alpha_c\!=\!1$.\label{fig2}}
\end{figure}

To numerically determine the transition point, we use the dc phase mobility, \(\displaystyle\mu\!\equiv\! \alpha/(2\pi)\lim_{\omega\rightarrow +0}\omega\langle\varphi\varphi\rangle_\omega\), that becomes zero (nonzero) in the SC (insulator) phase, where \(\langle\varphi\varphi\rangle_\omega\) is the Fourier transform of the ground-state phase correlation function \cite{Schmidt83,Kimura04,Schon90}. In the transformed frame, we can express it as
\eqn{
  \mu &= &\lim_{\omega\rightarrow +0} \sum_{n=0}^{\infty} \omega_{n0}\mu_{n0}\delta(\omega-\omega_{n0}),\label{mobility_exact}\\
  \mu_{n0}&\equiv &\alpha\bigl|\langle 0|\hat \Xi|n\rangle\bigr|^2, 
}
where \(\omega_{n0}\) is the  \(n\)-th excitation frequency, and we introduce the mobility matrix element \(\mu_{n0}\) with \(|n\rangle\) being the \(n\)-th energy eigenstate in the frame after the unitary transformation. 
We find that it suffices to calculate the dominant matrix element \(\mu_{10}\) for the purpose of locating the transition point.

Typical NRG flows of \(\mu_{10}\) in the bsG model are shown in Fig.~\ref{fig2}(a). As the energy scale is renormalized to lower regimes, the mobility eventually converges to zero in the SC phase \(\alpha\!>\!\alpha_c\), while it remains nonzero in the insulator phase \(\alpha\!<\!\alpha_c\). For each Wilson parameter $\Lambda$, we determine the critical value \(\alpha_c(\Lambda)\) by estimating the crossover scale \(N(\alpha)\) from NRG flows of \(\mu_{10}\) and assuming \(N(\alpha)\!\propto\!(\alpha-\alpha_c)^{-1}\). We then extrapolate the results to \(\Lambda\!\to\!1\) and locate the transition point \cite{Bulla03,*Bulla05}. As shown in Fig.~\ref{fig2}(b), our NRG results are consistent with the analytical value \(\alpha_c\!=\!1\) in the scaling limit \(\epsilon_J\!\equiv\!E_J/\hbar W \!\rightarrow \!0\).

Previous studies used the bsG model~\eqref{ham4} as a supposedly effective Hamiltonian of RSJ, which led to the vertical phase boundary at $\alpha_c\!=\!1$ (red dashed line in Fig.~\ref{fig1}(a)). The rationale behind this argument is that the capacitance term \(\nu\) is expected to be irrelevant from its scaling dimension and thus might be simply taken to be zero in Eq.~\eqref{ham3} while replacing UV cutoff by $\alpha E_C/(\pi\hbar)$ without affecting low-energy physics \cite{Fisher85,Affleck01}. 
However, the validity of this treatment must be carefully reexamined because the UV  theory~\eqref{ham3} possesses a large capacitance term $\nu\!\gg\!1$, and its low-energy theory may go beyond perturbative regimes during RG processes before reaching to a fixed point with $\nu\!=\!0$. To make concrete predictions, we thus need to resort to a nonperturbative analysis that consistently incorporates possible renormalization induced by the capacitance term $\nu$.

\begin{figure}[t]
    \includegraphics[width=8cm]{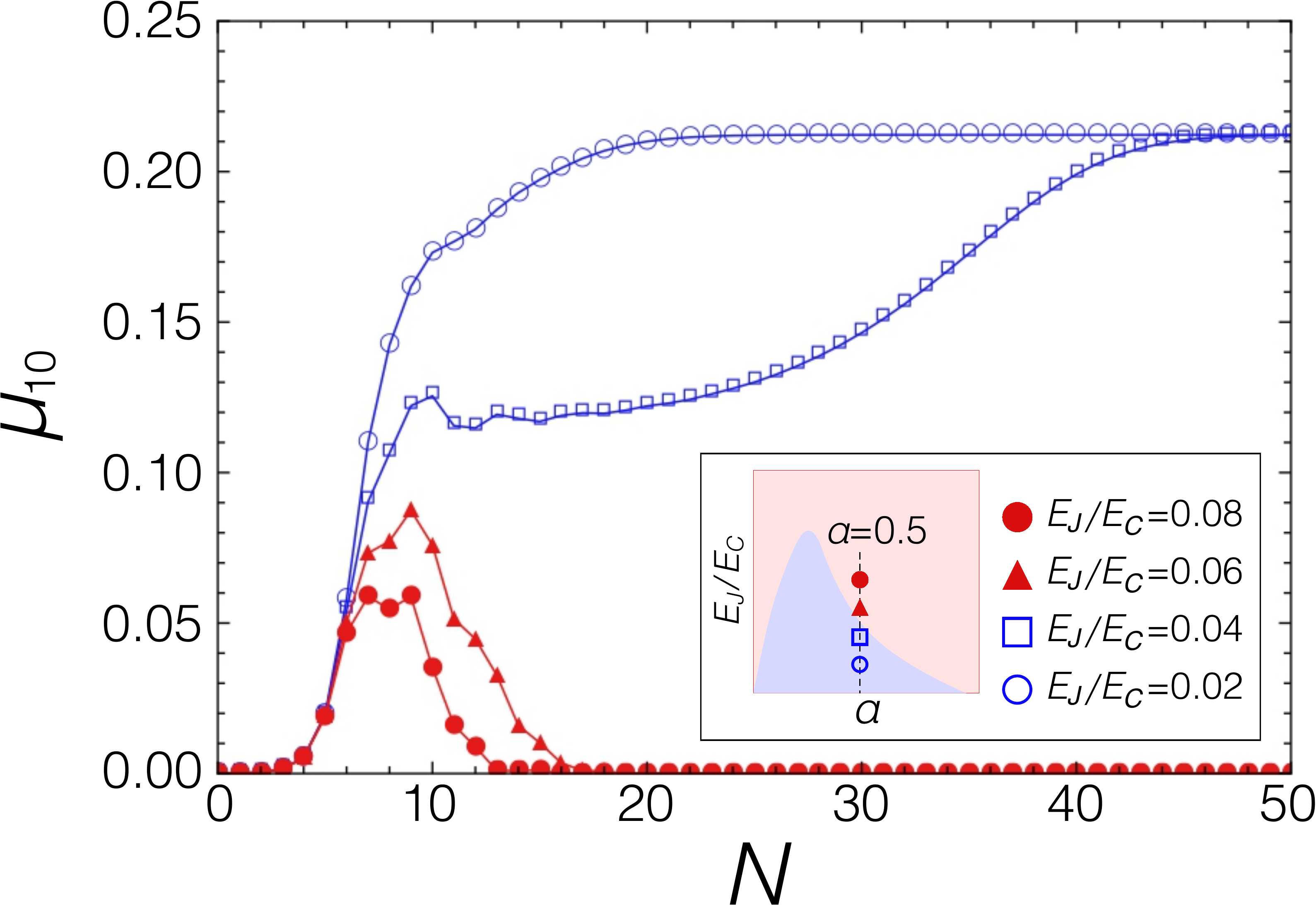}
    \caption{NRG flows of \(\mu_{10}\) in the exact RSJ Hamiltonian~\eqref{ham3} at different $E_J/E_C$. The inset indicates the corresponding parameter regions in the phase diagram. The system flows to the insulator fixed point with nonzero $\mu_{10}$ when \(E_J/E_C\) is sufficiently small (blue curves). In contrast, the system nonmonotonically flows to the SC fixed point with zero $\mu_{10}$ if \(E_J/E_C\) surpasses a critical value (red curves).  Parameters are \(\alpha \!=\! 0.5, \Lambda\!=\!2.0\), and \( \epsilon_C \!=\! 0.05\).
    \label{fig3}}
\end{figure}

{\it NRG analysis of the exact RSJ Hamiltonian.---}
To achieve this, we now apply the NRG approach to the exact RSJ Hamiltonian~\eqref{ham3}. To be concrete, we fix the charging energy \(\epsilon_C\!=\!E_C/(\hbar W) \!=\! 0.05\) and vary the Josephson coupling as \(0\!<\!E_J/E_C\!\lesssim\!0.4\), for which the dimensionless couplings satisfy the wideband condition $\epsilon_{J,C}\!\ll\!1$ at UV scale. We confirm that our NRG analysis is already converged against the wideband limit \cite{SM1}. 
Figure~\ref{fig3} shows typical NRG flows of \(\mu_{10}\) at $\alpha\!=\!0.5$. At the beginning of RG procedures, the mobility \(\mu_{10}\) always grows and the system tends to flow into the insulator phase. When \(E_J/E_C\) is sufficiently small, \(\mu_{10}\) keeps increasing and the system ultimately reaches to the insulator fixed point (blue curves in Fig.~\ref{fig3}). Surprisingly, when \(E_J/E_C\) surpasses a certain threshold value \((E_J/E_C)_c\), the mobility \(\mu_{10}\) turns from increasing to decreasing during RG processes and the system eventually flows to the SC fixed point (red curves in Fig.~\ref{fig3}). The convergence of these flows becomes slower as one gets closer to the transition point (e.g.,  \(E_J/E_C \!=\! 0.04\) in Fig.\ref{fig3}). We determine critical values \((E_J/E_C)_c\) shown in Fig.~\ref{fig1}(a) by extrapolating the Wilson parameter \(\Lambda\!\to\!1\) for each $\alpha$ \cite{SM1}.

Figure~\ref{fig4} shows fixed-point values of the phase coherence \(\langle \cos(\varphi)\rangle\) and the mobility \(\mu_{10}\) at different $\alpha$ and $E_J/E_C$. The phase coherence gives inductive contribution to supercurrent carried by the ground state \cite{Joyez13,Murami20,Safi11}. The behaviors of \(\langle \cos(\varphi)\rangle\) and \(\mu_{10}\) are consistent with each other; \(\langle \cos(\varphi)\rangle\) vanishes and \(\mu_{10}\) becomes nonzero in the insulator phase while the opposite is true in the superconducting phase. These results clearly indicate that the superconducting (insulating) phase at \(\alpha\!>\!0\) corresponds to the phase-localized (phase-delocalized) phase. It is also notable that both \(\langle\cos(\varphi)\rangle\) and \(\mu_{10}\) unambiguously indicate the reentrant transition from the insulator to SC phase as the resistance \(R\) is significantly increased beyond $R_Q$ (i.e., $\alpha\!\ll\!1$). In fact, in the limit \(R\!\to\!\infty\), JJ completely decouples from the environment and should remain superconducting (cf.~Eq.~\eqref{ham1}); our results in Fig.~\ref{fig4} Fig.~\ref{fig1}(a) are consistent with this expectation.  

\begin{figure}[t]
    \includegraphics[width=8cm]{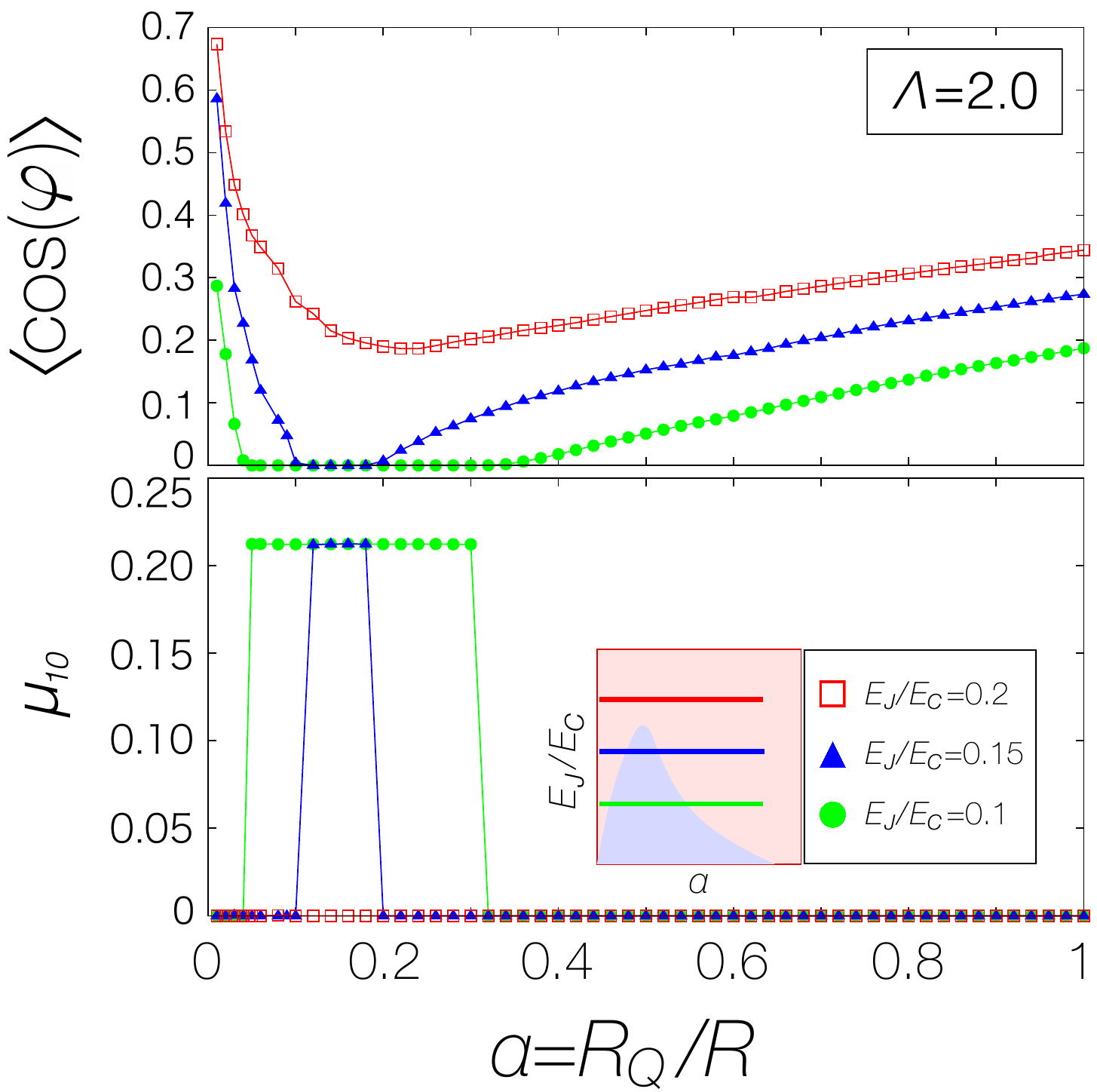}
    \caption{Phase coherence \(\langle\cos(\varphi)\rangle\)  and the mobility \(\mu_{10}\) plotted against \(\alpha=R_Q/R\). The inset indicates the corresponding parameter regions in the phase diagram at \(\Lambda\!=\!2.0\). At sufficiently large $E_J/E_C$, the system always resides in the SC phase (red curve). When $E_J/E_C$ is lower than a critical value $(E_J/E_C)_c$, there appears the insulating region as well as the reentrant transition into the SC phase in $\alpha\!\ll\!1$ (green and blue curves).  Parameters are \(\Lambda = 2.0\) and \(\epsilon_C = 0.02\).
    \label{fig4}}
\end{figure}

{\it FRG analysis.---} 
To understand these NRG results on a deeper level, we employ a nonperturbative analytical approach known as the FRG  \cite{WETTERICH199390,DUPUIS20211}. We use the functional ansatz retaining the most relevant Fourier mode, $\cos(\varphi)$, and  go beyond the local potential approximation by including the (field-independent) wavefunction renormalization, resulting in the following set of flow equations  \cite{SM1}:
\eqn{
d_{l}\ln\epsilon_{J}&=&1-\int_{0}^{\infty}\frac{dy}{\pi}g(y),\label{frgej}\\
d_{l}\ln\epsilon_{C}^{-1}&=&-1+\epsilon_{J}^{2}\int_{0}^{\infty}\frac{dy}{\pi}h(y),\label{frgec}
}
where $l\!=\!\ln(\Lambda_{0}/\Lambda)$ is the logarithmic RG scale, the dimensionless parameters satisfy $\epsilon_{J(C)}\!=\!E_{J(C)}/\Lambda_0\!\ll\! 1$ at UV scale $\Lambda\!=\!\Lambda_0$, and the integrals of $g,h$ give positive values~\cite{SM1}.

When $\epsilon_J\!\ll\!1$, the flow equation~\eqref{frgej} has the simple asymptotes depending on $\epsilon_C$,
\eqn{
d_{l}\ln\epsilon_{J}\overset{\epsilon_{J}\ll1}{\simeq}\begin{cases}
1-\frac{\sqrt{2\epsilon_{C}}}{8}>0 & \epsilon_{C}^{-1}\gg 1\\
1-\frac{1}{\alpha} & \epsilon_{C}^{-1}\to0
\end{cases}\label{frgec2},
}
the latter of which reproduces the well-known perturbative result implying the {\it presence} of DQPT at $\alpha_c\!=\!1$  \cite{Fisher85}. Notably, however, the former shows that the Josephson coupling $\epsilon_J$ is relevant at any $\alpha$ in UV regimes. This fact together with Eq.~\eqref{frgec} suggests that the supposedly irrelevant term $\nu\propto\epsilon_C^{-1}$ can significantly grow at low-energy scales due to the nonperturbative corrections, i.e., it can be dangerously irrelevant.

To determine fixed points the theory ultimately flow to, we numerically solve Eqs.~\eqref{frgej} and~\eqref{frgec}, and obtain the flow diagram in  Fig.~\ref{fig1}(b) \footnote{We remark that an apparent upper bound on the Josephson coupling at  $\epsilon_J=1$ in the flow diagram~\ref{fig1}(b) is an artifact due to our approximation which truncates (less relevant) higher-order Fourier modes $\cos(n\varphi)$ with $n\geq 2$.}. 
Due to the dangerously irrelevant term, when $E_J/E_C$ is larger than a critical value, the theory flows into the SC fixed point even when $\alpha\!<\!1$, leading to the {\it absence} of  DQPT in transmon regimes. The insulator phase is then  strongly suppressed to deep charge regimes $E_J/E_C\!\ll\!1$ with $\alpha\!<\!1$ \footnote{We confirm that the inclusion of higher Fourier modes does not affect our conclusions.}. 

At any $E_J/E_C$, the theory initially flows in favor of the insulator phase since the ratio obeys $d_l(\epsilon_J/\epsilon_C)\!<\!0$ in UV regimes $\epsilon_{J,C}\!\ll\!1$. At an intermediate low-energy scale, however, the theory enters nonperturbative regimes and can eventually exhibit the bifurcating flows to  different fixed points depending on $E_J/E_C$ (top panel in Fig.~\ref{fig1}(b)). This competition between renormalized Josephson and charging couplings  explains the nonmonotonic NRG flows found in Fig.~\ref{fig3}.

{\it Discussions.---}
The proposed phase boundary in Fig.~\ref{fig1}(a) is not vertical, which may appear to contradict with what is expected from the duality argument \cite{Bulgadaev84,Guinea85,Fisher85,Schmidt83}. The origin of this apparent inconsistency originates from the dangerously irrelevant term \(\nu\) discussed above. Indeed, only if \(\nu\) can be safely neglected, one can establish the duality between the weak and strong corrugation regimes \cite{Guinea85,SM1}.

In the strong corrugation regime \(E_J/E_C \!\gg\!1\),  it was argued \cite{Schmidt83,Guinea85} that the RSJ Hamiltonian can be approximated by the tight-binding model of phase localized states at \(\varphi \!=\! 2\pi\mathbb{Z}\). This model exhibits the transition at $\alpha_{c}\!=\!1$, which seems to be inconsistent with our results showing the absence of transition in transmon regimes. This apparent contradiction originates from a failure of the tight-binding description under the wideband condition $E_{C}\!\ll\!\hbar W$, in which a cutoff-dependent term invalidates the level truncation in each cosine well \cite{SM1}. 

Meanwhile, if one considers the opposite limit \(E_{C}\!\gg\!\hbar W\), both the tight-binding description and the duality argument are expected to be valid without such ambiguities. This parameter regime corresponds to the left sides of our FRG phase diagram (Fig.~\ref{fig1}(b)). Indeed, in this limit, our results are consistent with the previous results predicting the transition at $\alpha_{c}\!=\!1$ for any $E_J/E_C$.

To experimentally test our predictions,  one has to take account of the lowest transmission-line frequency \(\omega_{\rm min} \!=\! \pi v/L\) and finite temperature \(k_BT\), which effectively introduce an IR cutoff in RG flows. One needs to renormalize to a sufficiently low-energy scale to attain small \(\langle\cos(\varphi)\rangle\) close to a fixed-point value; this requires a sufficiently large system size and low temperature. For typical parameters of the insulator phase, \(\alpha\!=\! 0.3\) and \(E_J/E_C \!=\! 0.04\), one needs \(\hbar\omega_{\rm min},k_BT\! \lesssim\! 0.01E_C\) to attain \(\langle \cos \varphi\rangle\!\lesssim\! 10^{-2}\) \cite{SM1}. 
These conditions are within reach of recent experiments \cite{KR19,KR21,MHP19,LS19} which have realized galvanic coupling of JJ to a high-impedance long transmission line. In particular, Refs.~\cite{KR19,KR21} realize  \(E_C/h\!=\! 5.4\,{\rm GHz}\), \(\omega_{\rm min}/2\pi\! =\! 63\,{\rm MHz}\),  \(L\!\simeq\! 10\,{\rm mm}\), and UV cutoff  \(W/2\pi\!\simeq\! 20\,{\rm GHz}\) in superconducting waveguides, while $E_J$ is flux-tunable. These parameters correspond to \(\hbar\omega_{\rm min}/E_C\!\simeq\! 0.01\) and \(k_BT/E_C \!\simeq\! T/250\,{\rm mK}\). Thus, we expect that DQPT can be observed in this parameter region at millikelvin temperatures. We note that our estimation seems to be consistent with recent report of absence of DQPT \cite{Murami20}, on which we speculate that the experimental parameters \(E_C/h\! =\! 13\textrm{-}54\,{\rm GHz}\), \(L \!=\! 16\,\mu{\text m}\) lead to finite-size effects causing residual phase coherence \(\langle\cos(\varphi)\rangle\)\footnote{The resistor used in the experiment in Ref.~\cite{Murami20} is a chromic resistor, and one may make a modest estimate of its IR cutoff \(\sim v/L\) using the typical sound velocity of Chromium \(v \sim 6\times 10^3\) m/s.}.

In summary, we provided a comprehensive understanding of the dissipative quantum phase transition in a Josephson junction, which has been controversial for many years. We performed both numerical and analytical nonperturbative renormalization group analyses and obtained the phase diagram (Fig.~\ref{fig1}) in which the insulator phase is strongly suppressed to the deep charge regime while, in the transmon regime, the system remains  superconducting at any dissipation strengths. The origin of the failure of conventional understandings was traced to a previously overlooked dangerously irrelevant term which turns to be relevant in genuinely nonperturbative regimes. Physically, this renormalization behavior corresponds to the eventual decrease of charging energy at low energies, which ultimately results in the enhancement of \(E_J/E_C\) and the phase localization. Our analysis and understanding developed here can be applied to a variety of systems ranging from strongly interacting light-matter systems to electron-phonon problems. We hope that our work stimulates further studies in these directions.

\begin{acknowledgments}
We are grateful to Eugene Demler, Shunsuke Furukawa, Atac Imamoglu, Hosho Katsura, Naoto Nagaosa, Masahito Ueda, and Takeru Yokota for fruitful discussions. Y.A. acknowledges support from the Japan Society for the Promotion of Science through Grant Nos.~JP19K23424 and JP21K13859. 
M.O. is supported in part by MEXT/JSPS KAKENHI Grant
Nos.~JP17H06462 and JP19H01808, JST CREST Grant
No.~JPMJCR19T2.
\end{acknowledgments}

\bibliography{reference_masuki}

\widetext
\pagebreak
\begin{center}
\textbf{\large Supplementary Materials}
\end{center}

\renewcommand{\theequation}{S\arabic{equation}}
\renewcommand{\thefigure}{S\arabic{figure}}
\renewcommand{\bibnumfmt}[1]{[S#1]}
\setcounter{equation}{0}
\setcounter{figure}{0}

\subsection{Derivation of the Hamiltonian of the resistively shunted Josephson junction}
\begin{figure}[b]
    \includegraphics[width = 12cm]{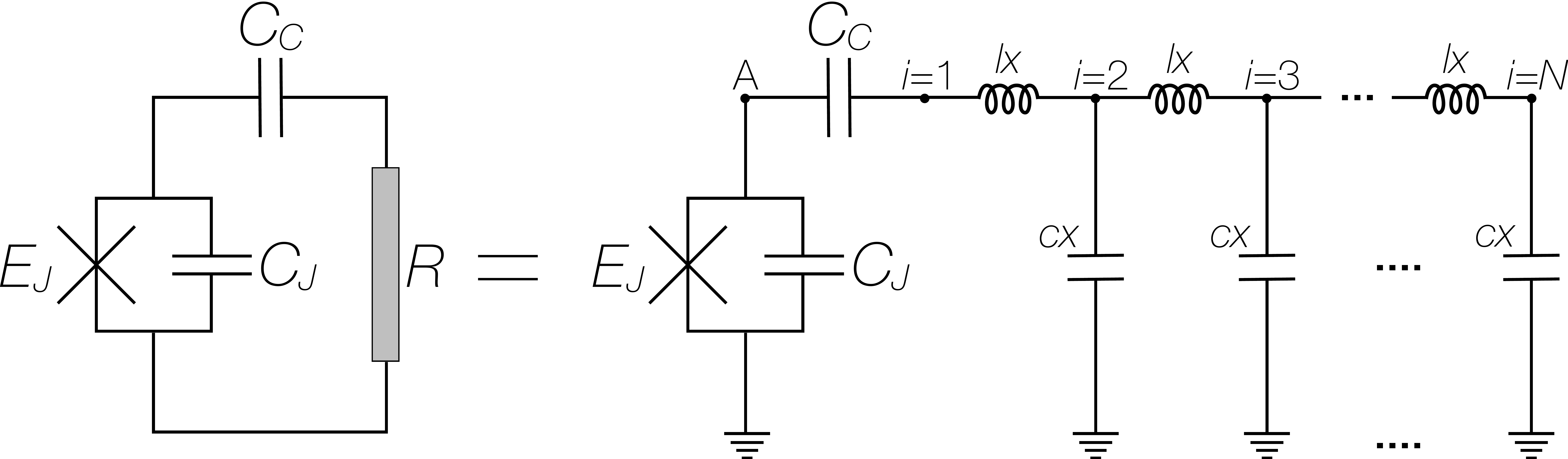}
    \caption{Microscopic model of the resistively and capacitively shunted Josephson junction. The resistor is represented by a lumped-element circuit with impedance \(R=\sqrt{l/c}\), propagation speed $v=1/\sqrt{lc}$, and length \(L=Nx\).   The Hamiltonian of the resistively shunted Josephson junction (Eq.~\eqref{ham2} in the main text) can be obtained by taking the limit \(C_C\rightarrow \infty\). \label{sm_fig1}}
\end{figure}

We here provide the derivation of the Hamiltonian~\eqref{ham2} in the main text from a microscopic point of view. The Josephson junction (JJ) shunted by a resistor can be modeled by a lumped-element circuit in Fig.~\ref{sm_fig1} (see also Refs.~\cite{GMF17,MM17,Leppakangas18}). In this microscopic model, the resistor is represented as a transmission line characterized by length \(L=Nx\) with integer $N$, propagation speed $v=1/\sqrt{lc}$, and impedance \(R=\sqrt{l/c}\), where \(c\) and \(l\) are the capacitance and inductance per unit length of the transmission line. We denote the coupling energy and the capacitance of the JJ as \(E_J\) and \(C_J\), respectively. Below we assume that \(N\) (or equivalently \(L\)) is taken sufficiently large; in the NRG analysis, we are interested in the thermodynamic limit \(L\rightarrow\infty\). For the sake of generality, we assume that the JJ and the resistor are capacitively coupled via the capacitance \(C_C\); we then obtain the Hamiltonian of the resistively shunted Josephson junction (RSJ) by taking the limit of \(C_C\rightarrow \infty\) at the end of the calculation.

The total Lagrangian of the lumped-element circuit in Fig.~\ref{sm_fig1} is given by
\begin{align}
  \mathcal L = \frac{C_J}{2}\dot{\Phi}^2 + E_J\cos\left(\frac{2e}{\hbar}\Phi\right) + \frac{C_C}{2}(\dot{\Phi}_1-\dot{\Phi})^2 + \sum_{i=2}^N \frac{cx}{2}\dot{\Phi_i}^2 - \sum_{i=2}^{N} \frac{(\Phi_i-\Phi_{i-1})^2}{2lx},
\end{align}
where \(\Phi_i(t) = \int_{-\infty}^t V_i(t')dt'\) is the flux given by the integral of the voltage at the \(i\)-th node, and \(\Phi(t)\) is the flux at the node A. We introduce the conjugate charge variables  \(Q_{i}\equiv \partial\mathcal L/\partial\dot{\Phi}_{i}\) and \(Q\equiv \partial\mathcal L/\partial\dot{\Phi}\), and quantize them via the commutation relations \([\hat\Phi_{i},\hat Q_{i}]  = i\hbar\) and  \([\hat\Phi,\hat Q]  = i\hbar\). After performing the Legendre transformation, we obtain the following Hamiltonian:
\begin{align}
  \hat H &= \frac{\left(\hat Q+\hat Q_1\right)^2}{2C_J} + \frac{\hat Q_1^2}{2C_C} + \sum_{i=2}^N \frac{\hat Q_i^2}{2cx} + \sum_{i=2}^{N} \frac{\left(\hat\Phi_i-\hat\Phi_{i-1}\right)}{2lx} -E_J\cos\left(\frac{2e}{\hbar}\hat\Phi\right)\\
  &=\frac{1}{2} \hat{\bm Q}^TA\hat{\bm Q} + \frac{1}{2lx}\hat{\bm \Phi}^TB\hat {\bm\Phi} -E_J\cos\left(\frac{2e}{\hbar}\hat\Phi\right)\label{sm_ham_hamstart},
\end{align}
where the vectors \(\hat{\bm Q}\) and \(\hat{\bm \Phi}\) represent
\eqn{
  \hat{\bm Q} &= (\hat Q,\hat Q_1,\ldots,\hat Q_N)^T,\\
  \hat{\bm \Phi} &= (\hat \Phi,\hat \Phi_1,\ldots,\hat \Phi_N)^T,\\
}
while \((N\!+\!1)\)-dimensional matrices \(A\) and \(B\) are defined by
\begin{align}
  A &= \left(\begin{array}{llllll}
    \frac{1}{C_J}&\frac{1}{C_J}&&&&\\
    \frac{1}{C_J}&\frac{1}{C_{p1}}&&&&\\
    &&\frac{1}{cx}&&&\\
    &&&\frac{1}{cx}&&\\
    &&&&\ddots&\\
    &&&&&\frac{1}{cx}
  \end{array}\right) \equiv \left(\begin{array}{cc}
    \frac{1}{C_J}&\bm w^T\\
    \bm w & \bar{A}
  \end{array}\right),\;\;\;
  B = \left(\begin{array}{cccccccc}
    0&&&&&&\\
    &1&-1&&&&\\
    &-1&2&-1&&&\\
    &&-1&2&\ddots&&\\
    &&&\ddots&\ddots&-1&\\
    &&&&-1&2&-1\\
    &&&&&-1&1
  \end{array}\right) \equiv \left(\begin{array}{cc}
    0&{\bf 0}\\
    {\bf 0} & \bar{B}
  \end{array}\right).
\end{align}
Here, \(C_{p1}= C_CC_J/(C_C+C_J)\) is the series capacitance seen by the resistor, \(\bar{A}\) and \(\bar{B}\) are \(N\)-dimensional matrices defined as the right-bottom part of \(A\) and \(B\), respectively, and \(\bm w\) is a \(N\)-dimensional vector defined by \(\bm w = (1/C_J,0,\ldots,0)^T\).

We next introduce a \((N\!+\!1)\)-dimensional regular matrix \(D\) in the following form:
\begin{align}
  D = \left(\begin{array}{cc}
    1 & 0\\
    0 & \bar{D}
  \end{array}\right),
\end{align}
and use it to rewrite the Hamiltonian~\eqref{sm_ham_hamstart} via the canonical transformation \(\hat{\bm \Phi} =D\hat{\bm \Psi}\) and \(\hat{\bm Q} = D^{-T}\hat{\bm P}\):
\begin{align}
  \hat H &= \frac{1}{2}\hat{\bm P}^TD^{-1}AD^{-T}\hat{\bm P} + \frac{1}{2lx}\hat{\bm \Psi}^T D^TBD \hat{\bm \Psi} - E_J\cos\left(\frac{2e}{\hbar}\hat \Psi\right),\label{sm_ham_after_canonical}
\end{align}
where
\eqn{
  D^{-1}AD^{-T} =\left(\begin{array}{cc}
    C_J^{-1} & \bm w^T\bar{D}^{-T}\\
    \bar{D}^{-1}\bm w&(\bar{D}^T\bar{A}^{-1}\bar{D})^{-1}
  \end{array}\right),\label{sm_ham_DAD}\;\;\;
  D^TBD = \left(\begin{array}{cc}
    0 & \\
    &\bar{D}^T\bar{B}\bar{D}
  \end{array}\right).
}
We can decouple \(\hat \Psi_i\) and \(\hat \Psi_j\) for \(i\neq j\) in Eq.~\eqref{sm_ham_after_canonical} by using a matrix \(\bar{D}\) that simultaneously diagonalizes \(\bar{D}^T\bar{A}^{-1}\bar{D}\) and \(\bar{D}^T\bar{B}\bar{D}\) in Eq.~\eqref{sm_ham_DAD}. One such a choice of \(\bar{D}\) is given by \(\bar{D} = \sqrt{C}\bar{A}^{1/2}O_B\), where \(C\) is a constant with the dimension of capacitance and \(O_B\) is an orthogonal matrix which diagonalizes the following symmetric matrix \(\bar{A}^{1/2}\bar{B}\bar{A}^{1/2}\): 
\begin{align}
  \bar{A}^{1/2}\bar{B}\bar{A}^{1/2} &= \frac{1}{cx}\left(\begin{array}{ccccc}
    \frac{cx}{C_{p1}}&-\!\left(\frac{cx}{C_{p1}}\right)^{1/2}&&&\\
    -\!\left(\frac{cx}{C_{p1}}\right)^{1/2}&2&-1&&\\
    &-1&\ddots&\ddots&\\
    &&\ddots&2&-1\\
    &&&-1&1
  \end{array}\right)
  = \frac{1}{cx}\left(\begin{array}{ccccc}
    \gamma^2&-\gamma&&&\\
    -\gamma&2&-1&&\\
    &-1&\ddots&\ddots&\\
    &&\ddots&2&-1\\
    &&&-1&1
  \end{array}\right),
\end{align}
where we introduce \(\gamma \equiv (cx/C_{p1})^{1/2}\). One can show that the matrix \(\bar{A}^{1/2}\bar{B}\bar{A}^{1/2}\) has an eigenvector \(\bm v=(v_1,\cdots,v_N)^T\) with eigenvalue \(\lambda = 4/(cx)\sin^2(\kappa/2)\). Each component of $\bm v$ and its eigenvalue are determined by the following relations
\eqn{
  (\bm{v})_m = \left\{
    \begin{array}{ll}
      \frac{1}{\gamma}\cos(\phi) &(m=1)\\\\
      \cos(\kappa (i-1+\phi)) & (m=2,3,\ldots,N)
    \end{array}
  \right.\label{sm_ham_cond_evec1}, \;\;\;\;\; \left\{
    \begin{array}{l}
      \cos(\phi) -\cos(\kappa +\phi) = \frac{4}{\gamma^2}\sin^2\left(\frac{\kappa }{2}\right)\cos(\phi)\\
      \cos(\kappa (N\!-\!1)+\phi) = \cos(\kappa N + \phi)
    \end{array}
  \right..
}
The latter conditions can be rewritten as
\begin{align}
  \left\{
    \begin{array}{l}
      \kappa(N\!-\!1/2) + \arctan\left(\frac{2C_{p1}-cx}{cx}\tan\left(\frac{\kappa}{2}\right)\right)\in \pi\mathbb{Z}\\
      \tan(\phi) = \frac{2C_{p1}-cx}{cx}\tan\left(\frac{\kappa}{2}\right)
    \end{array}
  \right..\label{sm_ham_cond_evec3}
\end{align}
In the continuous limit \(N\!\rightarrow\!\infty\), Eq.~\eqref{sm_ham_cond_evec3} can be approximately solved by
\begin{align}
  \kappa = \kappa_i \simeq \frac{i\pi}{N},\;\;\;\ \tan(\phi) \simeq \frac{2C_{p1}}{cx}\tan\left(\frac{\kappa_i}{2}\right)\label{sm_ham_solved_evec}
\end{align}
with \(i=0,1,\ldots,N\!-\!1\). We then denote the normalized corresponding eigenvector labeled by wavenumber \(\kappa = \kappa_i\) as \(\bm v_{\kappa_i}\). In particular, for \(i\neq 0\), the first component of \(\bm v_{\kappa_i}\) is given by
\begin{align}
  (\bm v_{\kappa_i})_1 \simeq \sqrt{\frac{2}{Ncx}}\sqrt\frac{C_{p1}}{1+\left(\frac{2C_{p1}}{cx}\tan\frac{\kappa_i}{2}\right)^2} = \sqrt{\frac{2}{Lc}}\sqrt\frac{C_{p1}}{1+\left(\frac{2C_{p1}}{cx}\tan\frac{\kappa_i}{2}\right)^2}.\label{sm_ham_1comp_of_evec}
\end{align}
From Eq.~\eqref{sm_ham_1comp_of_evec} and \((\bm v_{\kappa_i})_1=(O_B)_{1,i+1}\), we get 
\begin{align}
  (\bar{D}^{-1}\bm w)_{i+1} &= \left(\frac{1}{\sqrt{C}}O_B^T\bar{A}^{-1/2}\bm w\right)_{i+1}\simeq \frac{1}{\sqrt{C}}\frac{C_{p1}}{C_J}\sqrt{\frac{2}{Lc}}\sqrt\frac{1}{1+\left(\frac{2C_{p1}}{cx}\tan\frac{\kappa_i}{2}\right)^2}.\label{sm_ham_coupling_osc}
\end{align}
Similarly, for \(i=0\), we get
\begin{align}
  (\bar{D}^{-1}\bm w)_{1} \simeq \frac{1}{\sqrt{C}}\frac{C_{p1}}{C_J}\sqrt{\frac{1}{Lc}}.
\end{align}
Also, from the definition of \(\bar D\), it follows that 
\eqn{
  \bar{D}^{-1}\bar{A}\bar{D}^{-T} = \frac{1}{C}I_N,\label{sm_ham_DAD_bar}\;\;\;
  \bar{D}^T\bar{B}\bar{D} \simeq C\ {\rm diag}\left(\lambda_0,\lambda_1,\ldots,\lambda_{N-1}\right),
}
where we define \(\lambda_i\) as \(\lambda_i = 4/(cx)\sin^2(\kappa_i/2)\). From Eqs.~\eqref{sm_ham_after_canonical},~\eqref{sm_ham_DAD},~\eqref{sm_ham_coupling_osc}, and~\eqref{sm_ham_DAD_bar}, we finally obtain
\begin{align}
  \hat H \simeq & \frac{\hat P^2}{2C_J} - E_J\cos\left(\frac{2e}{\hbar}\hat \Psi\right) + \sum_{i=1}^{N-1}\left(\frac{\hat P_i^2}{2C} + \frac{C}{2}\frac{4\sin^2(\frac{\kappa_i}{2})}{lcx^2}\hat \Psi_i^2 \right) + \sum_{i=1}^{N-1} \hat P\hat P_i \frac{1}{\sqrt{C}}\frac{C_{p1}}{C_J}\sqrt{\frac{2}{Lc}}\sqrt\frac{1}{1+\left(\frac{2C_{p1}}{cx}\tan\frac{\kappa_i}{2}\right)^2}.\label{sm_ham_ham_obtained1}
\end{align}
Here, we omit the zero mode with \(\kappa_0\!=\!0\) because \(\hat P_0\) commutes with the total Hamiltonian and can be absorbed in the charge offset attenuated by the total circuit capacitance \cite{Houzet20}. The third term in the right hand side of Eq.~\eqref{sm_ham_ham_obtained1} represents a collection of environmental harmonic oscillators, while the last term represents the interaction between the JJ and environmental modes.

To rewrite the Hamiltonian, we introduce the annihilation and creation operators of the bosonic environmental modes, \(\hat b_i\) and \(\hat b_i^\dagger\) with \(i=1,2,\ldots,N\!-\!1\), by
\eqn{
  \hat \Psi_i = \sqrt{\frac{\hbar}{2C\omega_i}}i\left(\hat b_i^\dagger - \hat b_i\right),\label{sm_ham_Psi}\;\;\; \hat P_i = -\sqrt{\frac{\hbar C\omega_i}{2}}\left(\hat b_i^\dagger + \hat b_i\right),
}
where we define \(\omega_i\) as \(\omega_i\!=\!2v/x\sin(\kappa_i/2)\). 
Substituting Eq.~\eqref{sm_ham_Psi} into Eq.~\eqref{sm_ham_ham_obtained1} and defining \(E_C\!=\!2e^2/C_J,\; \hat N\!=\!\hat P/(2e),\; \varphi\!=\!2e\hat\Psi/\hbar\), and \(\alpha\!=\!R_Q/R\!=\!h/(4e^2 R)\), we obtain
\eqn{
  \hat H &\simeq & E_C\hat N^2 - E_J\cos\left(\varphi\right) + \sum_{i=1}^{N-1}\hbar\omega_i\hat b_i^\dagger\hat b_i -\hat N \sum_{i=1}^{N-1} \hbar g_{i}(\hat b_i + \hat b_i^\dagger),\label{sm_ham_rsj}\\
  g_{i}& =& \frac{C_{p1}}{C_J}\sqrt{\frac{2\pi v}{L\alpha}}\sqrt{\frac{\omega_i}{1+\left(\frac{2C_{p1}}{cx}\tan\frac{\kappa_i}{2}\right)^2}}.\label{sm_ham_ham_obtained2}
}
Equation~\eqref{sm_ham_ham_obtained2} shows that the capacitive coupling \(g_i\) is strongly suppressed in high-frequency oscillator modes. Therefore, we can approximate \(\tan(\kappa_i/2)\) and \(\sin(\kappa_i/2)\) by \(\sim\kappa_i/2\), leading to the simplified expressions:
\begin{align}
  g_i =& \frac{C_{p1}}{C_J}\sqrt{\frac{2\pi v}{L\alpha}}\sqrt{\frac{\omega_i}{1+\left(\frac{C_{p1}}{C_J}\frac{\pi\hbar\omega_i}{E_C\alpha}\right)^2}},\;\;\;\ \omega_i=\frac{i\pi v}{L}.
\end{align}
In the limit of \(C_C\rightarrow\infty\), the series capacitance \(C_{p1}\) reduces to \(C_J\) and thus Eq.~\eqref{sm_ham_rsj} reproduces the RSJ Hamiltonian~\eqref{ham2} in the main text.

\subsection{Details about the numerical renormalization group (NRG) analysis}
We here provide the full technical details about the NRG analysis performed in the present work. We start from the Hamiltonian after the unitary transformation \cite{YA21,*ashida21} (see Eq.~{\eqref{ham3}} in the main text):
\eqn{
  \hat H = -E_J\cos(\varphi+\hat \Xi) + \sum_k \hbar\omega_k\hat b_k^\dagger\hat b_k,\label{sm_nrg_ham1}
 }
 where we recall that
 \eqn{
  \hat \Xi = i\sum_k \frac{g_k}{\omega_k} \left(\hat b_k^\dagger - \hat b_k\right),}
  and
  \eqn{
  g_k = \sqrt{\frac{2\pi v}{\alpha L}\frac{\omega_k}{1+\left(\frac{\nu\omega_k}{W}\right)^2}},\;\; \omega_k = vk, \;\;\nu=\frac{\pi}{\alpha\epsilon_C},\;\; \epsilon_C = \frac{E_C}{\hbar W}.}
Here, the summation is taken over \(k\!=\!n\pi/L\) with \(n=1,2,\cdots, M\), and \(W\!=M\pi v/L\) is the frequency cutoff. Note that \(\varphi\) in Eq.~\eqref{sm_nrg_ham1} commutes with the Hamiltonian and  taken to be \(\varphi = 0\) in the following. Below we use the unit \(\hbar = W = 1\) while \(W\) is made explicit in equations when desirable.

To extend Wilson's NRG approach to the present problem setting, we first need to represent the environment as a collection of harmonic oscillators with continuous modes. To this end, we rewrite the Hamiltonian~\eqref{sm_nrg_ham1} by
\begin{align}
  \hat H =& -\epsilon_J\cos\left[i\sum_k \delta^{1/2} h(\omega_k) \left(\hat b_k^\dagger - \hat b_k\right)\right] + \sum_k\omega_k\hat b_k^\dagger\hat b_k,\label{sm_nrg_ham2}
\end{align}
where we define
\begin{align}
  &h(\omega) = \sqrt{\frac{2}{\alpha\omega}\frac{1}{1+\left(\nu\omega\right)^2}},\;\;\delta = \frac{\pi v}{L}, \;\;\epsilon_J = \frac{E_J}{\hbar W}. 
\end{align}
In the thermodynamic limit \(L\!\rightarrow\!\infty\), this Hamiltonian becomes equivalent to the following Hamiltonian with continuous environmental modes:
\begin{align}
  \hat H &= -\epsilon_J \cos\left(\hat \Xi\right) + \int_0^Wd\omega\ \omega \hat c_\omega^\dagger\hat c_\omega,\label{sm_nrg_ham3}\\
  \hat \Xi &= i\int_0^Wd\omega\  h(\omega) \left(\hat c_\omega^\dagger-\hat c_\omega\right),\label{sm_nrg_Xi}
\end{align}
where \(\hat c_\omega\) and \(\hat c_\omega^\dagger\) satisfy the commutation relation \([\hat c_\omega,\hat c_{\omega'}^\dagger] = \delta(\omega-\omega')\). To check this equivalence, we introduce a complete set of orthonormal functions \(\psi_{nm}\) on \([0,W]\) with \(n\in \mathbb{Z}_{\geq 0}\) and \(m\in\mathbb{Z}\) by
\begin{align}
  \psi_{nm}(\omega) = \left\{
    \begin{array}{cl}
       \delta^{-\frac{1}{2}}\exp\left(\frac{2\pi mi\omega}{\delta}\right) &\ \omega\in[n\delta,(n+1)\delta] = \left[\frac{n\pi v}{L},\frac{(n+1)\pi v}{L}\right]\\
       0 &\ \text{otherwise}
    \end{array}
  \right..
\end{align}
In the basis of \(\psi_{nm}\), the operators \(\hat c_\omega\) and \(\hat c_\omega^\dagger\) can be written as
\eqn{
  \hat c_\omega = \sum_{n,m} \psi_{nm}(\omega)\hat c_{nm},\label{sm_nrg_cmn}\;\;\;
  \hat c_\omega^\dagger = \sum_{n,m} \psi^*_{nm}(\omega)\hat c_{nm}^\dagger,
}
where \(\hat c_{nm}^{(\dagger)}\) is the annihilation (creation) operators of the mode \(\psi_{mn}\). They satisfy the commutation relation \([\hat c_{nm},\hat c_{n'm'}^\dagger] = \delta_{nn'}\delta_{mm'}\). Using Eq.~\eqref{sm_nrg_cmn}, we can rewrite Eq.~\eqref{sm_nrg_Xi} as
\begin{align}
 \hat \Xi = i\sum_{n,m} \left(\int_0^W d\omega\ h(\omega)\psi^*_{nm}(\omega)\right) \hat c_{nm} + \text{H.c.},\label{sm_nrg_Xidash}
\end{align}
where in the thermodynamic limit \(L\!\rightarrow\!\infty\) the integral gives zero unless \(m=0\), i.e., the modes with \(m\neq 0\) are decoupled from the system. We then rewrite each term of Eq.~\eqref{sm_nrg_ham3} as
\begin{align}
  \hat \Xi &= i\sum_{n\in \mathbb{Z}_{\geq 0}} \delta^{\frac{1}{2}} h(n\delta)(\hat c_{n0}^\dagger -\hat c_{n0}),\\
  \int_0^W d\omega\ \omega \hat c_\omega^\dagger\hat c_\omega &= \sum_{n\in \mathbb{Z}_{\geq 0}} n\delta \hat c_{n0}^\dagger\hat c_{n0} + \text{(the decoupled modes)}.
\end{align}
Thus, identifying \(\hat c_{n0}\) as \(\hat b_k\) in Eq.~\eqref{sm_nrg_ham2} and noting the fact that the summation in Eq.~\eqref{sm_nrg_ham2} is taken over \(\omega_k = n\pi v/L = n\delta\) with \(n\in\mathbb{Z}_{>0}\), the Hamiltonian~\eqref{sm_nrg_ham3} is indeed equivalent to  Eq.~\eqref{sm_nrg_ham2} in the thermodynamic limit.

We can further simplify the transformed Hamiltonian~\eqref{sm_nrg_ham2} by introducing a monotonically increasing function \(g : [0,g^{-1}(W)]\rightarrow [0,W],\ \varepsilon \mapsto \omega=g(\varepsilon)\) to change the variable as \(\omega\to\varepsilon\). We denote the corresponding annihilation (creation) operator \(\hat a_\varepsilon^{(\dagger)}\) by
\begin{align}
  \hat a_\varepsilon^{(\dagger)} = \sqrt{\frac{dg(\varepsilon)}{d\varepsilon}}\hat c_{g(\varepsilon)}^{(\dagger)},
\end{align}
which satisfies the  usual commutation relation \([\hat a_\varepsilon, \hat a_{\varepsilon'}^\dagger] = \delta(\varepsilon-\varepsilon')\). In terms of these operators, the Hamiltonian~\eqref{sm_nrg_ham3} is represented  as
\begin{align}
  \hat H &= -\epsilon_J\cos\left(i \int_0^{g^{-1}(W)} d\varepsilon\ s(\varepsilon)(\hat a_\varepsilon^\dagger-\hat a_\varepsilon)\right) + \int_0^{g^{-1}(W)} d\varepsilon\ g(\varepsilon)\hat a_\varepsilon^\dagger\hat a_\varepsilon,\label{sm_nrg_ham4}\\
  s(\varepsilon) &= \sqrt{\frac{dg(\varepsilon)}{d\varepsilon}} h(g(\varepsilon))\label{sm_nrg_s1}.
\end{align}
In this change of variables, we first choose \(g(\varepsilon)\) and it automatically determines  \(s(\varepsilon)\) via Eq.~\eqref{sm_nrg_s1}. Equivalently, it is also possible to first choose \(s(g^{-1}(\omega))\) as a function of \(\omega\) and determine \(g(\varepsilon)\) in such a way that Eq.~\eqref{sm_nrg_s1} is satisfied:
\begin{align}
  \frac{dg^{-1}(\omega)}{d\omega} = \left(\frac{h(\omega)}{s(g^{-1}(\omega))}\right)^2.\label{sm_nrg_s2}
\end{align}
In our calculations, we choose \(s(g^{-1}(\omega)) = h(\omega)^2\), for which Eq.~\eqref{sm_nrg_s2} leads to 
\begin{align}
  g^{-1}(\omega) = \frac{\alpha\omega^2}{4}\left(1+\frac{\nu^2\omega^2}{2}\right).\label{sm_nrg_ginverse}
\end{align}

To perform the NRG analysis, we logarithmically discretize the interval \([0,W]\) by \(\{d_n\!=\!W\Lambda^{-n}\}_{n\in\mathbb{Z}_{\geq 0}}\), where \(\Lambda\) is the Wilson pamameter. We also define \(e_n \equiv g^{-1}(d_n)\) and introduce a complete set of orghonormal functions \(f_{nm}\) on \([0,g^{-1}(W)]\) with \(n\in\mathbb{Z}_{\geq 0}\) and \(m\in\mathbb{Z}\) by
\begin{align}
  f_{nm}(\varepsilon) = \left\{
    \begin{array}{cl}
      \displaystyle \frac{1}{\sqrt{e_n-e_{n+1}}} \exp\left(\frac{2m\pi i\varepsilon}{e_n-e_{n+1}}\right) & \varepsilon \in [e_{n+1},e_n]\\
      0 & \text{otherwise}
    \end{array}
  \right..
\end{align}
In the basis of \(f_{nm}\), the operators \(\hat a_\varepsilon\) and \(\hat a_\varepsilon^\dagger\) can be expanded as
\eqn{
  \hat a_\varepsilon = \sum_{n,m} f_{nm}(\varepsilon)\hat a_{nm},\label{sm_nrg_aepsilon}\;\;\;
  \hat a_\varepsilon^\dagger = \sum_{n,m} f^*_{nm}(\varepsilon) \hat a_{nm}^\dagger,
}
where \(\hat a_{nm}^{(\dagger)}\) is the annihilation (creation) operator of the mode \(f_{nm}\). Using Eq.~\eqref{sm_nrg_aepsilon}, the integrals in Eq.~\eqref{sm_nrg_ham4} can be written as
\begin{align}
   i\int_0^{g^{-1}(W)} d\varepsilon s(\varepsilon)(\hat a_\varepsilon^\dagger -\hat a_\varepsilon) &= i\sum_{n,m} \hat a_{nm}^\dagger \int_{d_{n+1}}^{d_n} d\omega f^*_{nm}(g^{-1}(\omega)) + {\rm H.c.}\\
  &\simeq i\sum_{n} \frac{d_{n}-d_{n+1}}{\sqrt{e_n-e_{n+1}}} (\hat a_{n0}^\dagger -\hat a_{n0})\label{sm_nrg_log1} 
\end{align}
and
\begin{align}
  \int_0^{g^{-1}(W)} d\varepsilon g(\varepsilon) \hat a_\varepsilon^\dagger\hat a_\varepsilon &= \sum_{n,m,m'} \hat a_{nm}^\dagger\hat a_{nm'} \int_{e_{n+1}}^{e_n}d\varepsilon\ g(\varepsilon) f^*_{nm}(\varepsilon)f_{nm'}(\varepsilon)\\
  &\simeq \sum_{n,m} \hat a_{nm}^\dagger\hat a_{nm} \int_{e_{n+1}}^{e_n}d\varepsilon\ g(\varepsilon) f^*_{nm}(\varepsilon)f_{nm}(\varepsilon)\label{sm_nrg_log2}.
\end{align}
Here, we neglect \(m\!\neq\!0\) modes in Eq.~\eqref{sm_nrg_log1} as well as   couplings with \(m\!\neq\!m'\) in Eq.~\eqref{sm_nrg_log2} in the same spirit of Wilson's NRG \cite{Wilson75, Bulla05}. This treatment becomes exact in the limit of \(\Lambda\!\rightarrow\!1\), and hence we need to extrapolate \(\Lambda\!\rightarrow\! 1\) at the end of calculations to obtain accurate results. Within this approximation,  \(m\!\neq\!0\) modes in Eq.~\eqref{sm_nrg_log2} are decoupled from the system and thus can be omitted; for the sake of notational simplicity, hereafter we shall drop the \(m\!=\!0\) index in the operators \(\hat a_{nm=0}^{(\dagger)}\). 

\begin{figure}[b]
  \includegraphics[width=12cm]{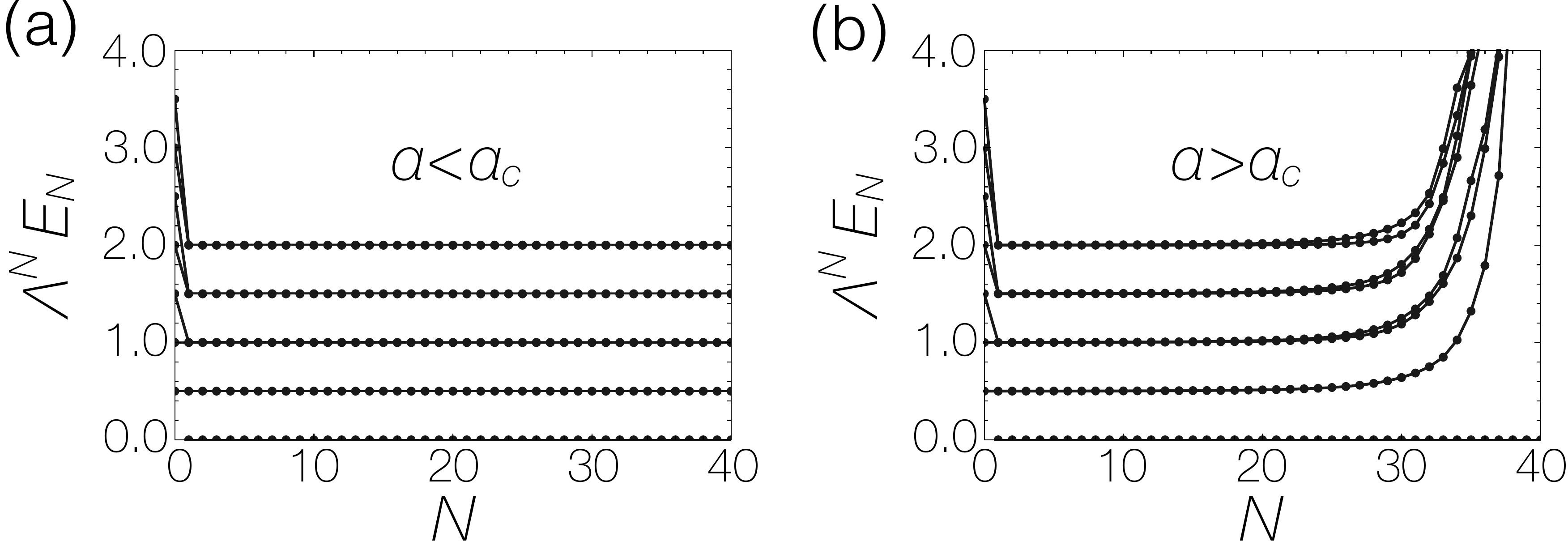}
  \caption{NRG flows of the excitation energies \(\Lambda^N E_N\) in the boundary sine-Gordon model (Eq.~\eqref{sm_nrg_ham_for_nrg1} with $\nu=0$) plotted against the number of RG steps \(N\). The spectrum remains almost the same when \(\alpha\!<\!\alpha_c\) (a), while excitation energies significantly increase when \(\alpha\!>\!\alpha_c\) (b). Parameters are \(\alpha = 0.9, \epsilon_J = 0.001, \Lambda = 2.0\) in (a) and \(\alpha = 1.2, \epsilon_J = 0.001, \Lambda = 2.0\) in (b).\label{sm_fig2}}
\end{figure}

From Eqs.~\eqref{sm_nrg_ham4},~\eqref{sm_nrg_ginverse},~\eqref{sm_nrg_log1} and~\eqref{sm_nrg_log2}, we finally arrive at the logarithmically discretized Hamiltonian:
\begin{align}
  \hat H &\simeq -\epsilon_J \cos(\hat \Xi) + \sum_{n=0}^{\infty} \gamma_n\hat a_n^\dagger\hat a_n,\label{sm_nrg_ham_for_nrg1}\\
  \hat \Xi &= i\sum_{n=0}^{\infty}\xi_n (\hat a_n^\dagger - \hat a_n),\label{sm_nrg_ham_for_nrg2}
  \end{align}
  where
  \eqn{
  \xi_n &= \sqrt{\frac{\Lambda-1}{\Lambda+1}}\sqrt{\frac{4}{\alpha}}\frac{1}{\sqrt{1+\frac{\nu^2}{2}\Lambda^{-2n}(1+\Lambda^{-2})}},\label{sm_nrg_ham_for_nrg3}\\
  \gamma_n &= 2\Lambda^{-n} \frac{\frac{1+\Lambda+\Lambda^{-2}}{3} + \nu^2\Lambda^{-2n}\frac{1+\Lambda^{-1}+\cdots+\Lambda^{-4}}{5} }{ (1+\Lambda^{-1})(1+\nu^2\Lambda^{-2n}\frac{1+\Lambda^{-2}}{2}) }.\label{sm_nrg_ham_for_nrg4}
}
We can now follow the usual procedure of the NRG calculations. Specifically, at the \(N\)-th step of NRG, we first diagonalize the Hamiltonian,
\begin{align}
  \hat H_N =& -\epsilon_J \cos(\hat \Xi_N) + \sum_{n=0}^{N} \gamma_n\hat a_n^\dagger\hat a_n,\label{sm_nrg_HN}\\
  \hat \Xi_N =&\ i\sum_{n=0}^N \xi_n (\hat a_n^\dagger - \hat a_n),
\end{align}
by retaining \(n_S\) lowest-energy eigenstates while truncating the higher excited states. We then add the \((N\!+\!1)\)-th bosonic mode, $\hat{a}_{N+1}$, and prepare the matrix elements of \(\hat H_{N+1}\). For this purpose, it is in practice useful to use the following relation:
\begin{align}
  \cos\left(\hat \Xi_{N+1}\right) = \frac{1}{2}\exp\left(i\hat\Xi_{N}\right)\exp\Bigl(\xi_{N+1}\left(\hat a_{N+1}-\hat a_{N+1}^\dagger\right)\Bigr) + \text{H.c.},\label{sm_nrg_cos}
\end{align}
where we prepare the matrix of \(\xi_{N+1}(\hat a_{N+1}\!-\!\hat a_{N+1}^\dagger)\) and exponentiate it to obtain the matrix elements of the operator $  \cos\left(\hat \Xi_{N+1}\right)$. In adding the \((N\!+\!1)\)-th bosonic mode, we include \(n_B\) lowest eigenstates of the number operator \(\hat a_{N\!+\!1}^\dagger\hat a_{N\!+\!1}\) : \(|0\rangle,|1\rangle,\ldots,|n_B\!-\!1\rangle\), in the similar manner as previously done for the spin-boson model \cite{Bulla05}. In the present work, we set \(n_S\!=\!50\) and \(n_B\!=\!300\) for the \(N\!=\!0\) mode while \(n_B\!=\!15\) for the other modes with \(N\geq 1\). 

We note that one can use the Bogoliubov transformation to change the Hamiltonian~\eqref{sm_nrg_ham_for_nrg1} into (in analogy with conventional NRG methods) either the chain-NRG form:
\begin{align}
  \hat H_{\rm chain} = -\epsilon_J\cos\Bigl(i\xi'_0(\hat b_0^\dagger-\hat b_0)\Bigr) + \sum_{n=0}^\infty \gamma'_n\hat b_n^\dagger\hat b_n + \sum_{n=1}^{\infty}\xi'_n\left(\hat b_{n-1}^\dagger-\hat b_{n-1}^\dagger\right)\left(\hat b_n^\dagger-\hat b_n^\dagger\right),\label{sm_nrg_chainNRG}
\end{align}
or the star-NRG form:
\begin{align}
  \hat H_{\rm star} = -\epsilon_J\cos\Bigl(i\xi''_0(\hat b_0^\dagger-\hat b_0)\Bigr) + \sum_{n=0}^\infty \gamma''_n\hat b_n^\dagger\hat b_n + \sum_{n=1}^{\infty}\xi'_n\left(\hat b_{0}^\dagger-\hat b_{0}^\dagger\right)\left(\hat b_n^\dagger-\hat b_n^\dagger\right),\label{sm_nrg_starNRG}
\end{align}
where \(\hat b_n^{(\dagger)}\) is the bosonic annihilation (creation) operator and \(\xi'_n,\gamma'_n,\xi''_n,\gamma''_n\) are constants of order \(\Lambda^{-n}\). However, comparing NRG results of these Hamiltonians~\eqref{sm_nrg_ham_for_nrg1},~\eqref{sm_nrg_chainNRG} and~\eqref{sm_nrg_starNRG}, we find that NRG analysis based on the Hamiltonian~\eqref{sm_nrg_ham_for_nrg1} consistently shows the best convergence and provides reliable results; in the other two cases, accumulations of errors due to truncations of higher Fock states are much more appreciable. All the numerical results in this paper are thus obtained by performing NRG to the Hamiltonian~\eqref{sm_nrg_ham_for_nrg1}.

As discussed in the main text, we first benchmark our NRG analysis by applying it to the boundary  sine-Gordon (bsG) model which corresponds to  taking the limit \(\nu\to 0\)  in Eq.~\eqref{sm_nrg_ham_for_nrg1}. 
Figure~\ref{sm_fig2} shows the typical NRG flows of the excitation energies, where \(E_N\) represents an energy eigenvalue of the Hamiltonian~\eqref{sm_nrg_HN} and is rescaled by the factor \(\Lambda^N\) in the plots. When the theory is renormalized to low-energy  regimes, the excitation spectrum almost remains the same for \(\alpha < 1\) (insulator phase), while excitation energies significantly increase for \(\alpha>1\) (superconducting phase). This is consistent with the interpretation that \(E_J\) is irrelevant in \(\alpha<1\) and relevant in \(\alpha>1\) (cf. Eq.~\eqref{sm_nrg_HN}). 
The corresponding flows of the mobility matrix element \(\mu_{10}\) are shown in Fig.~\ref{fig2}(a) in the main text. We estimate the crossover scale \(N(\alpha)\) by determining \(N\) at which \(\mu_{10}\) becomes half of the insulator fixed-point value (for instance, in Fig.~\ref{fig2}(a), the fixed-point value of \(\mu_{10}\) in the insulating phase is around \(0.21\)). From the flow equation of \(\epsilon_J\) (cf. the latter of Eq.~\eqref{frgec2} in the main text), we expect the scaling \(N(\alpha)\propto 1/(\alpha-\alpha_c)\), which is then used to determine \(\alpha_c\) by fitting the values of $N(\alpha)$. We explain this estimation of \(\alpha_c\) in more detail in the section below.

\begin{figure}
  \includegraphics[width = 17cm]{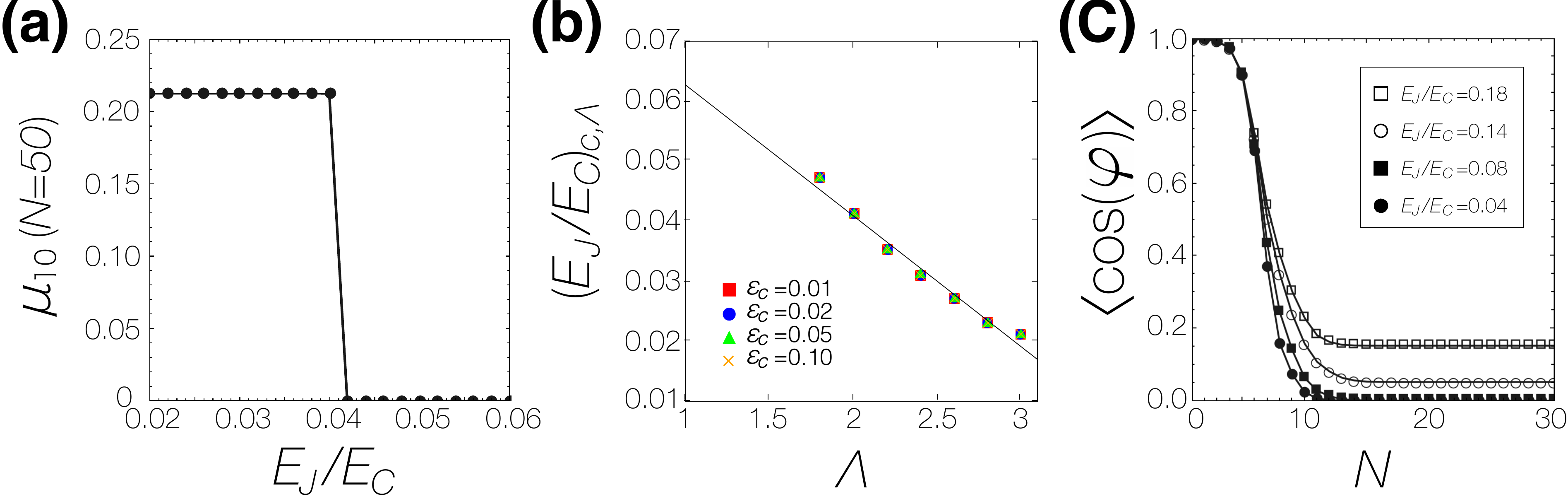}
  \caption{NRG results of the RSJ Hamiltonian (Eq.~\eqref{sm_nrg_ham_for_nrg1} with $\nu>0$). (a) Fixed-point values of the mobility matrix element \(\mu_{10}\) plotted against \(E_J/E_C\). The results are obtained after performing a sufficiently large number of RG steps \(N=50\). The value of \(\mu_{10}\) changes abruptly at a critical value \((E_J/E_C)_{c,\Lambda}\) as \(E_J/E_C\) is increased. Parameters are \(\alpha = 0.5, \Lambda = 2.0\), and \(\epsilon_C = 0.05\). (b) Extrapolation of the critical values \((E_J/E_C)_{c,\Lambda}\) to the Wilson parameter \(\Lambda\rightarrow 1\). Parameter is \(\alpha = 0.5\). (c) Flows of the phase coherence \(\langle\cos(\varphi)\rangle\) plotted against the number of RG steps \(N\). Parameters are \(\alpha = 0.3, \Lambda =2.0\), and \(\epsilon_C = 0.05\).\label{sm_fig3}}
\end{figure}

Figure~\ref{sm_fig3}(a) shows the fixed-point values of \(\mu_{10}\) in the RSJ Hamiltonian (Eq.~\eqref{sm_nrg_ham_for_nrg1} including the $\nu$-term), which are obtained after a sufficiently large number of RG steps.
As $E_J/E_C$ is increased, one can see that a fixed-point value of \(\mu_{10}\) abruptly changes from nonzero to zero at a critical value \((E_J/E_C)_{c,\Lambda}\). To determine the transition point  \((E_J/E_C)_{c}\) shown in Fig.~\ref{fig1}(a) in the main text, we extrapolate the Wilson parameter \(\Lambda\rightarrow 1\) for these results \((E_J/E_C)_{c,\Lambda}\) at each $\alpha$; see Fig.~\ref{sm_fig3}(b).
It is clear that there is no appreciable change in \((E_J/E_C)_{c,\Lambda}\) when varying \(\epsilon_C = E_C/W\), which implies that the cutoff \(W\) is taken sufficiently large and the calculation has already converged in the wideband limit \(W\gg E_C, E_J\).
Figure~\ref{sm_fig3}(c) shows typical NRG flows of the phase coherence \(\langle\cos(\varphi)\rangle\). While the phase coherence converges to a nonzero value in the superconducting phase (open rectangles/circles), it exponentially converges to zero in the insulating phase (filled rectangles/circles). 
These results show that this superconducting (insulating) phase at \(\alpha\!>\!0\) can be interpreted as the phase-localized (phase-delocalized) phase as also mentioned in the main text.
We also remark that, in the current implementation, NRG calculations become technically difficult when \(\alpha\!\ll\!1\) for which \(\nu\) is singularly large at UV scale. In particular, it is challenging to extrapolate the Wilson parameter \(\Lambda\!\rightarrow\!1\) and to quantitatively locate the transition point in this case. Nevertheless, we expect that the phase boundary in Fig.~\ref{fig1}(a) is qualitatively accurate even in this regime; for instance, the numerical results at a fixed Wilson parameter \(\Lambda\) (e.g., Fig.~\ref{fig4}) unambiguously indicate the reentrant transition occurring in the deep charge regime. 

\subsection{Estimation of the critical point of the boundary sine-Gordon model}
\begin{figure}
  \includegraphics[width = 11cm]{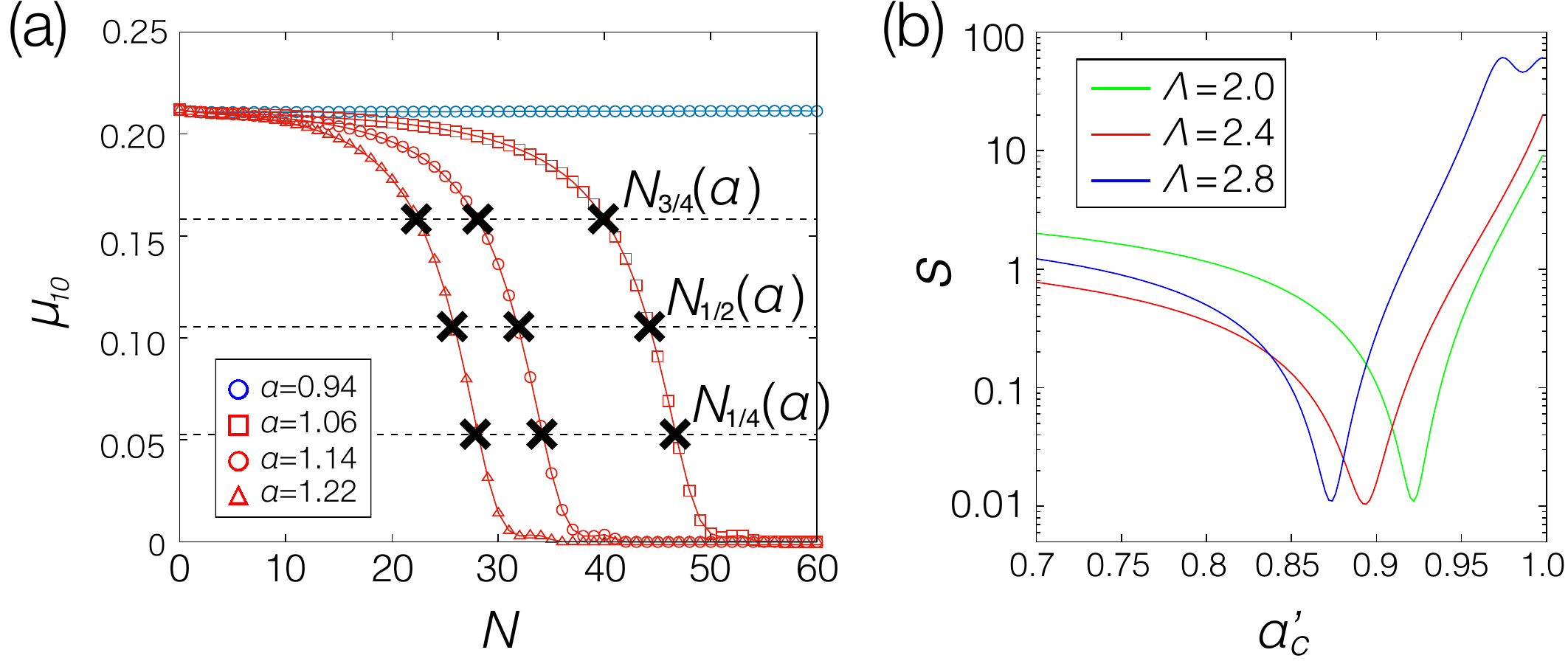}
  \caption{(a) The estimation of the crossover scale \(N(\alpha)\) of the bsG model. The parameters are \(\epsilon_J = 0.001, \Lambda = 2.0.\) (b) The typical behavior of the standard error of estimates \(s\) as a function of \(\alpha_c'\). We determine \(\alpha_{c,\Lambda}\) as the \(\alpha_c'\) which gives the smallest standard error of estimates \(s\). The parameter is \(\epsilon_J = 0.001\).\label{sm_fig4}}
\end{figure}

We here provide the detailed explanation of the estimation of the critical point  \(\alpha_c\) in the boundary sine-Gordon (bsG) model. First, as mentioned in the main text, we have estimated the crossover scale \(N(\alpha)\) as the number of RG steps \(N\) at which the mobility matrix element \(\mu_{10}\) becomes half of the insulator-fixed point value. To examine possible dependence on a choice of the threshold value, we have also estimated \(N(\alpha)\) at which \(\mu_{10}\) becomes one quarter and three quarters of the insulator fixed-point value. We denote these \(N(\alpha)\) as \(N_{1/2}(\alpha), N_{1/4}(\alpha)\), and \(N_{3/4}(\alpha)\), respectively. We show these estimated crossover scales \(N(\alpha)\) in Fig.~\ref{sm_fig4}(a). To be concrete, we complement the value of \(\mu_{10}\) at \(N=0\) by using its insulator fixed-point value. We then estimate the crossover scale \(N(\alpha)\) by linearly interpolating the values of \(\mu_{10}\) for integer \(N\). The estimation error in \(N_{1/2}(\alpha), N_{1/4}(\alpha)\), and \(N_{3/4}(\alpha)\) was \(\sim 0.1\) in this procedure.

Second, we vary \(\alpha_c'\) and fit \(N_x(\alpha)\)\((x=1/4, 1/2, 3/4)\) against \((\alpha-\alpha_c')^{-1}\) by a straight line. We calculate the standard error of estimates \(s\) in this fitting for each \(\alpha_c'\), and we determine \(\alpha_{c,\Lambda}\) as the \(\alpha_c'\) which gives the smallest standard error of estimates \(s\). Typical behaviors of \(s\) as a function of \(\alpha_c'\) are shown in Fig.~\ref{sm_fig4}(b). From this graph, we estimate the error of \(\alpha_{c,\Lambda}\) as 0.02.

\begin{figure}
  \includegraphics[width = 11cm]{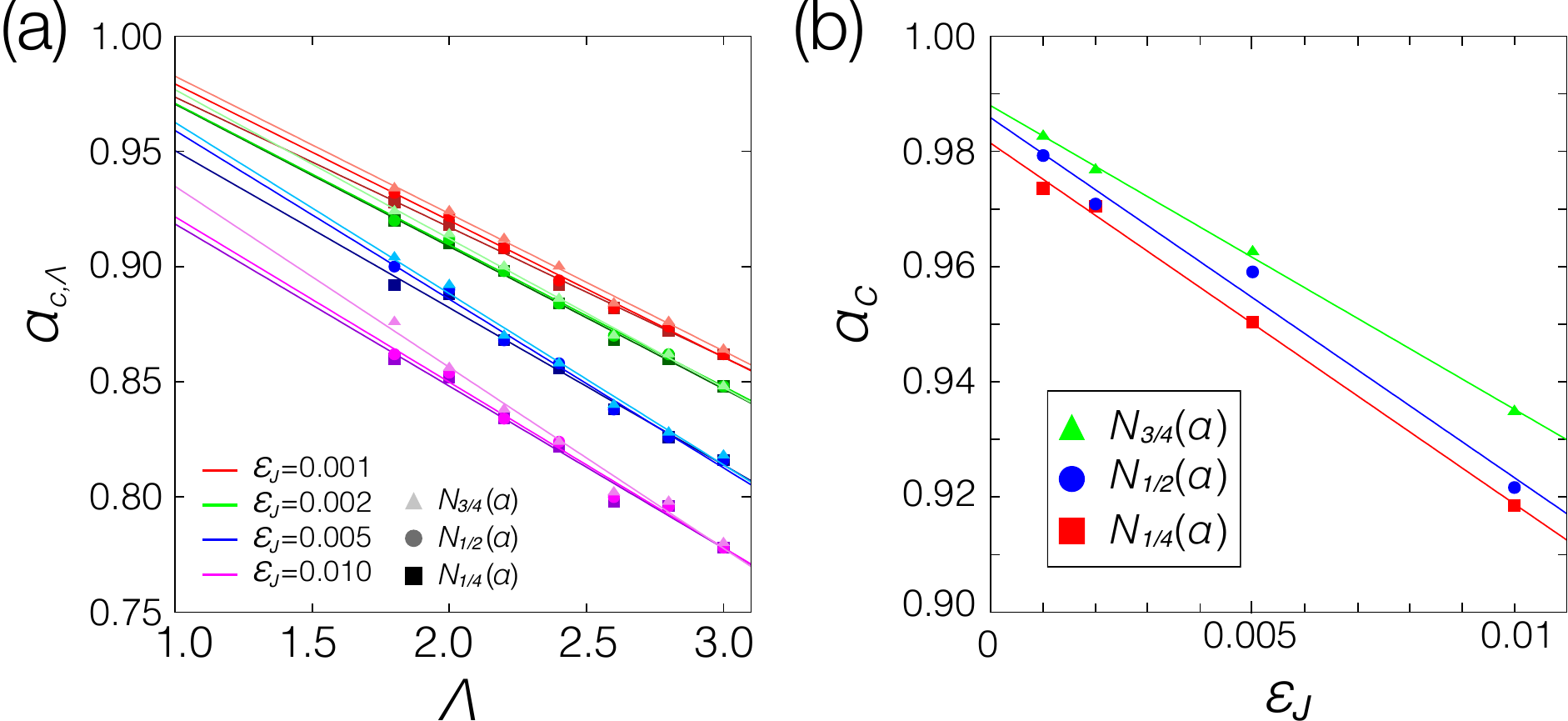}
  \caption{(a) Extrapolations of the critical point value \(\alpha_{c,\Lambda}\) to the Wilson parameter \(\Lambda\!\to\!1\). (b) Extrapolations of the critical point value \(\alpha_{c}\) to the scaling limit \(\epsilon_J\!\to\!1\).\label{sm_fig5}}
\end{figure}

Third, we extrapolate the Wilson parameter \(\Lambda\!\to\!1\) and decide \(\alpha_{c,\Lambda}\) for each \(\epsilon_J = E_J/(\hbar W)\). Finally, we extrapolate \(\epsilon_J\!\to\!0\) and estimate \(\alpha_c\) in the bsG model. These extrapolations are shown in Figs.~\ref{sm_fig5}(a) and (b) respectively. We can see that the data are well fitted by straight lines. It can also be inferred from the plot that the change in \(\alpha_c\) when using \(N_{1/2}(\alpha), N_{1/4}(\alpha)\), and \(N_{3/4}(\alpha)\) is sufficiently small. Combining these errors and the error in deciding \(\alpha_{c,\Lambda}\) in the second step explained above, we estimate the critical point value of the bsG model \(\alpha_c\) as 0.99(2). 

\subsection{Relation with earlier results}
We briefly discuss how the present results are related to some of the results obtained before (e.g., Refs.~\cite{Schmidt83,Fisher85,Guinea85,Affleck01}). Below we conclude that all of these previous analyses, including the duality argument and analysis based on the tight-binding description in the strong corrugation regimes, are unambiguously valid only when the charging energy satisfies $E_C\gg \hbar W$, while in general break down under the wideband condition $E_C\ll \hbar W$ which is considered in our work and actually feasible in realistic experimental systems as discussed in the main text (see e.g., Refs.~\cite{KR19,KR21}). We note that our FRG analysis also supports this conclusion since the previous results are consistently reproduced in the limit $\nu\propto \hbar W/E_C\to 0$, which correspond to the left sides of Fig.~\ref{fig1}(b) in the main text. We set \(\hbar = v = 1\) below in this section.
 
\subsubsection{Duality argument}
We shall first reexamine the validity of the duality argument. To this end, we rewrite the  Hamiltonian~\eqref{ham2} in the main text as 
\begin{align}
  \hat H &= -E_J\cos(\varphi) - \hat N\sum_k g_k(\hat b_k^\dagger+\hat b_k) + \sum_k k\hat b_k^\dagger\hat b_k + \sum_k \frac{g_k^2}{k}\hat N^2,\label{sm_previous_ham1}
\end{align}
where we use the sum rule \(E_C = \sum_k g_k^2/k\) with
\begin{align}
  g_k &= \sqrt{\frac{2\pi}{\alpha L}\frac{k}{1+\left(\frac{\nu k}{W}\right)^2}}.
\end{align}
Since the capacitance term $\nu\propto W/E_C$ is expected to be irrelevant from its scaling dimension, previous studies simplified the analysis in the weak corrugation regime $E_J/E_C\ll1$ by essentially taking the limit \(\nu \rightarrow 0\) in the Hamiltonian~\eqref{sm_previous_ham1} \cite{Fisher85,Affleck01}, leading to  
\begin{align}
  \hat H^{\nu\rightarrow 0} = -\frac{E_J}{2}\sum_{n\in\mathbb{Z}}\left(|n\rangle\langle n+1| + {\rm H.c.} \right)- \hat N\sqrt{\frac{2\pi}{\alpha}}\sum_k \sqrt{\frac{k}{L}}\left(\hat b_k^\dagger+\hat b_k\right) + \sum_k k \hat b_k^\dagger\hat b_k + \sum_k \frac{2\pi}{\alpha L}\hat N^2\label{sm_previous_ham2}, 
\end{align}
where \(|n\rangle\) is an eigenstate of the charge operator \(\hat N\). 

Meanwhile, previous studies argued that the strong corrugation regime $E_J/E_C\gg 1$ can be analyzed on the basis of the following effective tight-binding model (see also discussions in the next subsection) \cite{Guinea85}:
\begin{align}
  \hat H^{E_J/E_C\gg 1} &= -\Delta \sum_{n\in\mathbb{Z}}\left(\hat c_{n+1}^\dagger\hat c_n + \hat c_n^\dagger\hat c_{n+1}\right) - \hat q \sqrt{2\pi\alpha}\sum_k\sqrt{\frac{k}{L}}\left(\hat b_k^\dagger+\hat b_k\right) + \sum_k k\hat b_k^\dagger\hat b_k + \sum_k \frac{2\pi\alpha}{L}\hat q^2,\label{sm_previous_Guinea}\\
  \hat q &= \sum_{n\in\mathbb{Z}} n\hat c_n^\dagger\hat c_n,
\end{align}
where $\hat{c}_n$ ($\hat{c}^\dagger_n$) represents the annihilation (creation) operator of a particle localized in the potential well labeled by $n$, and $\Delta\propto e^{-\sqrt{32E_J/E_C}}$ is the hopping matrix element between the adjacent localized states.  
One can easily check that these two Hamiltonians are equivalent; this suggests the duality between the weak \eqref{sm_previous_ham2} and strong \eqref{sm_previous_Guinea} corrugation regimes via the correspondence: \(E_J/2\!\leftrightarrow\!\Delta\) and \(1/\alpha\!\leftrightarrow\!\alpha\). As discussed in the main text, these Hamiltonians can be mapped to the boundary sine-Gordon model after a unitary transformation and its perturbative renormalization group analysis indicates the phase transition at \(\alpha = 1\).  Then, the duality between weak \(E_J/E_C\ll 1\) and strong \(E_J/E_C\gg 1\) corrugation regimes seems to suggest that the transition occurs at $\alpha=1$ for any \(E_J/E_C\). This results in the vertical phase boundary as shown in the red dashed line in Fig.~\ref{fig1}(a) in the main text. 

However, as demonstrated in the main text, these arguments in general break down when we consider the setup satisfying the wideband condition \(E_{C}\ll W\) as appropriate for experimental systems~\cite{KR19,KR21}. The crucial point in this case is that the capacitance term $\nu\propto W/E_C$ takes a large value at UV scale and thus can be dangerously irrelevant in RG sense, i.e., it can turn into relevant in nonperturbative regimes. Thus, even in the weak corrugation regime, it cannot be justified to neglect the capacitance term $\nu$ as done in deriving Eq.~\eqref{sm_previous_ham2} from  Eq.~\eqref{sm_previous_ham1}. Said differently, the Hamiltonian~\eqref{sm_previous_ham2} is no longer a faithful description to analyze the low-energy properties of the RSJ. We will see below that the tight-binding description in the strong corrugation regime also becomes invalid  under the wideband condition. These lead to the breakdown of the duality argument above and explain the reason why the obtained phase boundary is not vertical, in contrast to what is expected from the duality argument (Fig.~\ref{fig1}(a) in the main text).

\subsubsection{Tight-binding or instanton description in the strong corrugation regime}
Next, we examine the validity of the tight-binding description which was introduced as an effective model in the strong corrugation regime $E_J/E_C\gg 1$ \cite{Guinea85}. To this end, we transform the original RSJ Hamiltonian~\eqref{ham1} in the main text by using the unitary transformation \(\hat V = \exp(i\hat n_r\varphi)\) (which is also known as the PZW transformation for obtaining the dipole-gauge Hamiltonian in the field of quantum optics):
\begin{align}
  \hat{\tilde{H}} \equiv \hat V^\dagger \hat H\hat V= E_C\hat N^2 - E_J\cos(\varphi) + \sum_k \frac{\alpha}{2\pi L}\varphi^2 + \varphi\sqrt{\frac{\alpha}{2\pi}}\sum_k\sqrt{\frac{k}{L}} i(\hat a_k^\dagger -\hat a_k) + \sum_k k\hat a_k^\dagger\hat a_k.\label{sm_previous_ham3}
\end{align}
In the strong corrugation regime \(E_J/E_C\gg1\), it was argued that higher bands in the JJ Hamiltonian \(\hat H_{\rm JJ}\!=\!E_C\hat N^2\!-\!E_J\cos(\varphi)\) can be truncated so that it can be  replaced by the (single-band) tight-binding model of phase localized Wannier states \(|n\rangle\) at \(\varphi\!=\!2\pi\mathbb{Z}\):
\begin{align}
  \hat H_{\rm JJ} \sim -\Delta\sum_{n\in\mathbb{Z}}\Bigl( |n\rangle\langle n+1| + |n+1\rangle\langle n| \Bigr),\label{sm_previous_Hs1}
\end{align}
where \(\Delta\propto e^{-\sqrt{32E_J/E_C}}\) is the hopping matrix element between the adjacent localized states. This also leads to the replacement of
 the JJ phase by
\eqn{
\varphi\sim 2\pi\hat{q}\;\;\;{\rm with}\;\;\;\hat{q}=\sum_{n\in\mathbb{Z}} n|n\rangle\langle n|.
} 
One may then approximate the original Hamiltonian~\eqref{sm_previous_ham3}  in the strong corrugation regime by the following tight-binding Hamiltonian \(\hat{\tilde{H}}\): 
\begin{align}
  \hat{\tilde{H}} \sim -\Delta\sum_{n\in\mathbb{Z}} \bigl( |n\rangle\langle n+1| + |n+1\rangle\langle n| \bigr) + \sum_k \frac{2\pi\alpha}{L}\hat{q}^2 + \hat{q}\sqrt{2\pi\alpha}\sum_k\sqrt{\frac{k}{L}} i(\hat a_k^\dagger -\hat a_k) + \sum_k k\hat a_k^\dagger\hat a_k,\label{sm_previous_ham4}
\end{align}
which gives Eq.~\eqref{sm_previous_Guinea}  with the identification $-i\hat{a}_k^\dagger\to\hat{b}_k^\dagger$. 
As noted above, this model~\eqref{sm_previous_ham4} is equivalent to the boundary sine-Gordon model and thus exhibits the phase transition at \(\alpha\!=\!1\) in $E_J/E_C\gg 1$. 

\begin{figure}[t]
    \includegraphics[width = 8cm]{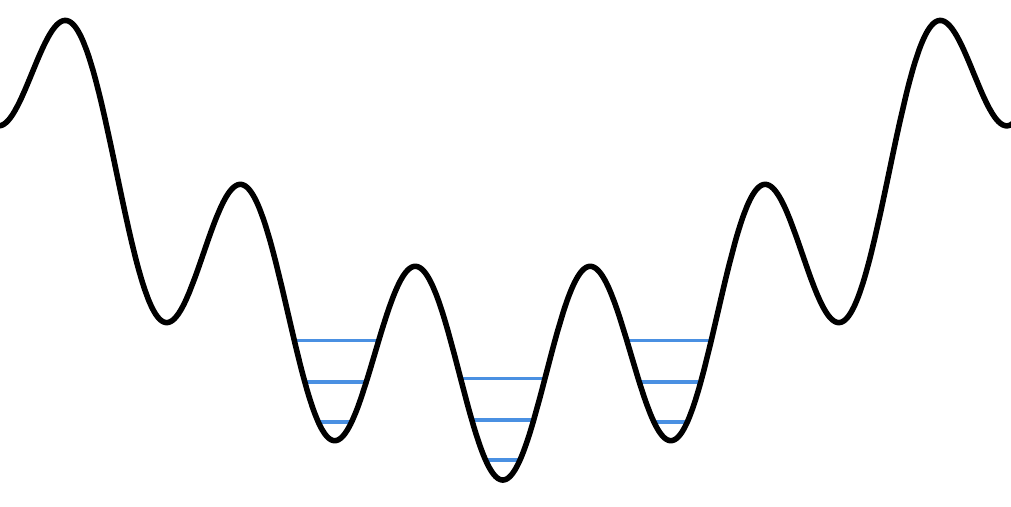}
    \caption{Potential profile in the ``dipole gauge" Hamiltonian~\eqref{sm_previous_ham3}, which consists of the Josephson term $-E_J\cos(\varphi)$ and a harmonic potential $\frac{\alpha W}{2\pi^2}\varphi^2$. In the latter, the potential strength is characterized by the UV cutoff $W$. In each well, there is a higher-excitation mode with the band gap $\sim \sqrt{E_JE_C}$. 
    The tight-binding description is expected to break down in the wideband limit $E_C\ll W$, where the energy acquisition $\sim W$ by hopping to the adjacent upper well surpasses the band gap $\sim \sqrt{E_CE_J}$ of JJ, and the truncation of higher bands becomes no longer valid.\label{sm_fig6}}
\end{figure}

We now point out that these arguments can be unambiguously justified only in the limit $E_C\!\gg\! W$, while again break down under the wideband condition $E_C\ll W$. Although this point seemed to be not made clear in the literature, such failure should originate from the breakdown of the tight-binding approximation made in deriving  Eq.~\eqref{sm_previous_ham4} from Eq.~\eqref{sm_previous_ham3}. Specifically, in the wideband limit \(E_{C}\ll W\) as considered in the present work, the harmonic potential term $\frac{\alpha W}{2\pi^2}\varphi^2$ in the Hamiltonian (the third term in the most right hand side of Eq.~\eqref{sm_previous_ham3})  becomes important and the energy acquisition $\sim W$ by hopping to the adjacent upper site eventually surpasses the band gap $\sim\sqrt{E_JE_C}$ of JJ (see Fig.~\ref{sm_fig6}).
At least in such a case, there seems to be no way to justify the truncation of higher bands and thus the tight-binding description is expected to be invalid. 
It is noteworthy that  such \(\varphi^2\)-term (often called ``counterterm") also appears in the spin-boson model \cite{Leggett87}, for which we expect  breakdown of the two-level treatment of a quantum particle in double-well potential under the wideband condition.

Meanwhile, if we consider the opposite limit \(E_C\!\gg\!W\), the \(\varphi^2\)-term is much smaller than $E_C\hat{N}^2$ and $-E_J\cos(\varphi)$, and thus might be simply neglected and the tight-binding description can still be valid. Another way to see this is to invoke the instanton analysis \cite{Schmidt83,NN99} which should be unambiguously valid when \(E_C\!\gg\!W\) and $E_J/E_C\gg 1$, for which the tunnelings probability to the adjacent localized states is very small even under the influence of dissipation. Thus, in the narrowband limit  \(E_C\!\gg\!W\), we expect that the phase transition occurs at \(\alpha = 1\). Indeed, this conclusion is consistent with our FRG analysis shown in Fig.~\ref{fig1}(b), where the system flows to the superconducting (insulating) fixed point for \(\alpha>1\)  (\(\alpha<1\)) in the limit \(1/\epsilon_C\to 0\), which corresponds to the left side of the flow diagrams.

\subsection{Details about the functional renormalization group (FRG) analysis}
We here provide the full technical details about the FRG analysis performed in the present work. Integrating out the bosonic environmental degrees of freedom, the imaginary-time action of the RCSJ at the zero temperature is given by (we set $\hbar\!=\!v\!=1$ in this section):
\eqn{
S[\varphi]=\int_{-\infty}^{\infty}d\tau\left[\frac{1}{4E_{C}}(d_{\tau}\varphi)^{2}-E_{J}\cos(\varphi(\tau))\right]+\frac{\alpha}{8\pi^{2}}\int_{-\infty}^{\infty}d\tau\int_{-\infty}^{\infty}d\tau'\left[\frac{\varphi(\tau)-\varphi(\tau')}{\tau-\tau'}\right]^{2}.
}
After applying the Fourier transformation to the first and last terms, we can rewrite the action as 
\eqn{
S[\varphi]=\frac{1}{2}\int_{-K}^{K}\frac{d\omega}{2\pi}\left(\frac{\alpha|\omega|}{2\pi}+\frac{\omega^{2}}{2E_{C}}\right)|\tilde{\varphi}(\omega)|^{2}-E_{J}\int_{-\infty}^{\infty}d\tau\cos(\varphi(\tau)),
}
where we denote the field variable in the frequency basis by $\tilde{\varphi}(\omega)=\int_{-\infty}^{\infty}d\tau\varphi(\tau)e^{-i\omega\tau}$. To perform the nonperturbative RG analysis, we consider the following functional ansatz for the effective action at an energy scale $\Lambda$:
\eqn{\label{frgansatz}
\Gamma_{\Lambda}[\varphi]=\frac{1}{2}\int_{-\infty}^{\infty}\frac{d\omega}{2\pi}\left(\frac{\alpha|\omega|}{2\pi}+\epsilon_{C}^{-1}\frac{\omega^{2}}{2\Lambda}\right)|\tilde{\varphi}(\omega)|^{2}-\epsilon_{J}\Lambda\int_{-\infty}^{\infty} d\tau\cos(\varphi(\tau)).
}
Here, for the sake of notational simplicity, we use the variable $\Lambda$ to represent RG energy scale (but not the Wilson parameter), and we introduce the field-independent dimensionless parameters, $\epsilon_J(\Lambda)$ and $\epsilon_C(\Lambda)$, which satisfy the following relations at a UV scale $\Lambda=\Lambda_0$
\eqn{
\epsilon_{C}(\Lambda_0)\equiv\frac{E_{C}}{\Lambda_0},\;\;\epsilon_{J}(\Lambda_0)=\frac{E_{J}}{\Lambda_0}.
}
The functional ansatz~\eqref{frgansatz} goes beyond the local potential approximation in the sense that the wavefunction renormalization is included, while we neglect a possible field dependence of the parameters and also truncate the less relevant, higher-order Fourier components $\cos(n\varphi)$ with $n\geq 2$. We confirm that the inclusion of these higher modes does not affect the results discussed in the main text. We can show that $\alpha$ will not be renormalized at all order at least in the present setup.

The flow equation follows from the Wetterich equation  \cite{WETTERICH199390,DUPUIS20211}:
\eqn{
\Lambda\partial_{\Lambda}\Gamma_{\Lambda}=\frac{1}{2}{\rm Tr}\left[\partial_{\Lambda}R_{\Lambda}G_\Lambda\right],\label{wetteq}
}
where $R_\Lambda$ is the regulator and the dimensionless propagator $G_\Lambda$ is defined by
\eqn{
G_\Lambda(\omega)\equiv\frac{1}{\epsilon_{C}^{-1}\frac{\omega^{2}}{2\Lambda^{2}}+\frac{\alpha|\omega|}{2\pi\Lambda}+\epsilon_{J}\cos(\varphi)+R_{\Lambda}/\Lambda}.
}
We can obtain the flow equation for the potential term $\epsilon_J$ by projecting the right hand side of Eq.~\eqref{wetteq} onto the most relevant Fourier mode:
 \eqn{
(1+\Lambda d_{\Lambda})\epsilon_{J}=\int_{0}^{2\pi}\frac{d\varphi}{2\pi}\cos\varphi\int_{-\infty}^{\infty}\frac{d(\omega/\Lambda)}{2\pi}G_{\Lambda}(\omega)\partial_{\Lambda}R_{\Lambda}.
 }
 To be concrete, we choose the simple regulator $R_{\Lambda}=\Lambda$ from now on, which allows us to analytically perform the integration over $\varphi$, leading to
\eqn{
d_{l}\ln\epsilon_{J}&=&1-\int_{0}^{\infty}\frac{dy}{\pi}g(y),\label{ejeq}\\
g(y)&\equiv&\frac{1}{\epsilon_{J}^{2}}\left[\frac{1+\frac{\alpha y}{2\pi}+\frac{\epsilon_{C}^{-1}y^{2}}{2}}{\sqrt{\left(1+\frac{\alpha y}{2\pi}+\frac{\epsilon_{C}^{-1}y^{2}}{2}\right)^{2}-\epsilon_{J}^{2}}}-1\right],
}
where $l=\ln(\Lambda_0/\Lambda)$ is the logarithmic RG scale. This gives Eq.~(\ref{frgej}) in the main text. In the perturbative limit $\epsilon_{J}\ll1$, we can simplify the flow equation~\eqref{ejeq} to
\eqn{
d_{l}\ln\epsilon_{J}\simeq1-\frac{F\left(\frac{8\pi^{2}}{\epsilon_{C}\alpha^{2}}\right)}{\alpha}\;\;\;{\rm with}\;\;\;F(t)\equiv\frac{\sqrt{1-t}-\frac{t}{2}\ln\left(\frac{1+\sqrt{1-t}}{1-\sqrt{1-t}}\right)}{\left(1-t\right)^{3/2}}.
}
Here, $F(t)$ is a monotonically decreasing function in $t\geq 0$ and satisfies $F(t)\simeq\frac{\pi}{2\sqrt{t}}$ at $t\gg1$ and $F(0)=1$, which lead to the asymptotic expressions in Eq.~(\ref{frgec2}) in the main text. 

To obtain the flow equation for $\epsilon_C^{-1}$, we take the second-order derivative of Eq.~\eqref{wetteq} with respect to the field variable and project it onto the lowest Fourier mode. The result is
\eqn{
\partial_{\Lambda}(\epsilon_{C}^{-1}/(2\Lambda))=\int_{0}^{2\pi}\frac{d\varphi}{2\pi}\left(\epsilon_{J}\sin(\varphi)\right)^{2}\int_{-\infty}^{\infty}\frac{d(\omega/\Lambda)}{2\pi}\partial_{\Lambda}R_{\Lambda}\left[G_{\Lambda}(\omega)\right]^{2}\lim_{\omega'\to0}\frac{G_{\Lambda}(\omega+\omega')-G_{\Lambda}(\omega)}{(\omega'/\Lambda)^{2}}.
}
Using the regulator $R_{\Lambda}=\Lambda$ and performing the integration over $\varphi$, we arrive at the flow equation as follows:
\eqn{
d_{l}\ln\epsilon_{C}^{-1}&=&-1+\epsilon_{J}^{2}\int_{0}^{\infty}\frac{dy}{\pi}h(y),\\
h(y)&=&\frac{1}{32\pi^{4}z}\frac{1}{\left[\left(1+zy^{2}+\frac{\alpha y}{2\pi}\right)^{2}-u^{2}\right]^{7/2}}\nonumber\\
&&\;\;\;\;\;\times\Bigl\{ \alpha^{4}y^{2}+2\pi\alpha^{3}y\left(2+5zy^{2}\right)+\pi^{2}\alpha^{2}\left[u^{2}+4(1+2zy^{2})(1+5zy^{2})\right]\nonumber\\
&&\;\;\;\;\;+8\pi^{3}\alpha yz\left[2u^{2}+(1+zy^{2})(1+9zy^{2})\right]+16\pi^{4}z(zy^{2}+2zy^{2}u^{2}+3z^{3}y^{6}+u^{2}+5y^{4}z^{2}-1)\Bigr\},
} 
where we abbreviate the variables as $u=\epsilon_{J}$ and $z=\epsilon_{C}^{-1}/2$ for the sake of notational simplicity. This equation provides Eq.~\eqref{frgec} in the main text.

\end{document}